\documentclass[12pt]{article}
\usepackage{jheppub}
\pdfoutput = 1

\usepackage{amsmath}
\usepackage{amssymb}
\usepackage{bbm}
\usepackage{bbold}
\usepackage{braket}
\usepackage{caption}
\usepackage{color}
    \definecolor{darkgreen}{rgb}{0,0.5,0}
    \definecolor{darkred}{rgb}{0.5,0,0}
    \definecolor{darkblue}{rgb}{0,0,0.6}
    \definecolor{purple}{rgb}{0.4,.2,0.7}

\usepackage{float}
\usepackage{graphicx}
\usepackage{hyperref}

\usepackage{mathabx}
\usepackage{mathtools}
\usepackage{pdfsync}
\usepackage{slashed}
\usepackage[normalem]{ulem}
\usepackage{upgreek}
\usepackage{url}

\usepackage[ragged]{footmisc}
\def\be{\begin{equation}}
\def\ee{\end{equation}}

\renewcommand{\tilde}{\widetilde}

\numberwithin{equation}{section}

\begin{document}
\title{Extremal black holes that are not extremal: maximal warm holes}
\author[a]{\'Oscar~J.C.~Dias,}
\author[b]{Gary~T.~Horowitz,}
\author[c]{Jorge~E.~Santos}
\affiliation[a]{STAG research centre and Mathematical Sciences, University of Southampton, UK}
\affiliation[b]{Department of Physics, University of California at Santa Barbara, Santa Barbara, CA 93106, U.S.A.}
\affiliation[c]{Department of Applied Mathematics and Theoretical Physics, University of Cambridge, Wilberforce Road, Cambridge, CB3 0WA, UK}

\emailAdd{O.J.Campos-Dias@soton.ac.uk}
\emailAdd{horowitz@ucsb.edu}
\emailAdd{jss55@cam.ac.uk}

\abstract{We study a family of four-dimensional, asymptotically flat, charged black holes that develop (charged) scalar hair as one increases their charge at fixed mass.  Surprisingly, the maximum charge for given mass is a nonsingular hairy black hole with nonzero Hawking temperature. The implications for Hawking evaporation are discussed.}

\maketitle
\section{Introduction}

In four-dimensional general relativity, there is a maximum charge and angular momentum that can be added to a black hole of given mass. In Einstein-Maxwell theory, these extremal black holes are characterized by having a degenerate horizon with zero Hawking temperature. In theories that also have (real) scalars fields exponentially coupled to  the Maxwell field, such as supergravity or string theory, the extremal limit is either singular \cite{Garfinkle:1990qj}, or similar to Einstein-Maxwell due to an attractor mechanism \cite{Andrianopoli:2006ub}.

In this paper, we show that there are black holes with a  different type of extremal limit. In the theory we consider, black holes again have a maximum charge for given mass\footnote{For simplicity, we will restrict our attention to static black holes with no rotation.}, but the extremal black hole can have a nondegenerate horizon with nonzero Hawking temperature. Our theory will include a scalar field, but unlike some of the theories mentioned above, the usual Reissner-Nordstr\"om (RN) solution (describing static charged black holes in Einstein-Maxwell theory) remains a solution in the theory with scalar added.

 It has recently been shown that if a massless scalar field is appropriately coupled to $F^2\equiv F_{ab}F^{ab}$, RN can become unstable to forming scalar hair, \emph{i.e.}, static scalar fields outside the horizon \cite{Herdeiro:2018wub,Fernandes:2019rez}.  This is because $F^2 < 0$ for an electrically charged black hole, and acts like a negative potential for the scalar near the horizon. When the charge is large enough, this destabilizes the scalar field and causes it to become nonzero.
 
We add a massive, charged scalar field $\psi$ to Einstein-Maxwell with a simple $|\psi|^2 F^2$ coupling. As before, when the electric charge is large enough, RN becomes unstable and develops charged scalar hair. Our original motivation for exploring this model was that its solutions are asymptotically flat analogs of the asymptotically anti-de Sitter (AdS) solutions known as holographic superconductors \cite{Gubser:2008px,Hartnoll:2008vx,Hartnoll:2008kx}, which have been extensively studied. In AdS, the charged scalar condenses at low temperature without any explicit coupling between the scalar and Maxwell field. Without the cosmological constant, however, this does not happen \cite{Hod:2015hza} and one needs to add a coupling like  $|\psi|^2 F^2$.  (There are also hairy black holes without this coupling if the charged scalar has an appropriate potential \cite{Herdeiro:2020xmb,Hong:2020miv}, but they do not branch off from RN.\footnote{In higher dimensions, there are examples of hairy black holes in  Einstein-Gauss-Bonnet gravity with a minimally coupled charged scalar field with no self interactions \cite{Grandi:2017zgz}.})

It was shown in \cite{Hartnoll:2020fhc} that the dynamics inside the horizon of a holographic superconductor is quite intricate. We study the dynamics inside the horizon of this asymptotically flat analog in a companion paper \cite{dhs}.
Here we focus on the solutions outside the horizon, and look at their extremal limit.
 As seen before \cite{Garfinkle:1990qj,Herdeiro:2018wub}, the hairy black holes can exceed the usual extremal limit and have $Q^2 > M^2$. However, unlike previous examples, we find that for some range of parameters, the maximum charge solution for fixed mass is a nonsingular hairy black hole with nonzero Hawking temperature. So although it is ``extremal" in the sense of having maximum charge, it is not a familiar  ``extremal black hole" with either  zero temperature or a singular horizon. We will call this new type of extremal black hole a ``maximal warm hole".
 
The existence of maximal warm holes raises puzzling questions about the endpoint of Hawking radiation. If a black hole continues to radiate neutral gravitons when it reaches its extremal limit, it would appear to create a naked singularity. Unlike the standard Planck mass naked singularity expected at the endpoint of the  evaporation of a neutral black hole, this could create a naked singularity with large mass. We will argue that this does not occur. 
 
 Our theory also contains charged solitons, and for completeness, we include a discussion of them. We find that they have a minimum mass, so they do \emph{not} exist arbitrarily close to Minkowski space. They also cannot be viewed as the limit of the hairy black holes as the black hole radius goes to zero.


\section{Equations of motion}

We start with the action
\begin{equation}\label{eq:action}
S= \int \mathrm{d}^{4}x \sqrt{-g}\left[R- F^2-4(\mathcal{D}_a\psi)(\mathcal{D}^a \psi)^\dagger-4 m^2 |\psi|^2-4 \alpha F^2 |\psi|^2\right]\,,
\end{equation}
where $\mathcal{D}=\nabla-i\,q\,A$ and $F=\mathrm{d}A$\,. This theory satisfies all the usual energy conditions  if the coupling constant $\alpha$ is positive, which we will assume is the case.

The equations of motion for this general action read
\begin{subequations}\label{EOM:S}
\begin{multline}
R_{ab}-\frac{R}{2}g_{ab}=2\left(1+4 \alpha |\psi|^2\right)\left(F_{ac}F_b^{\phantom{b}c}-\frac{g_{ab}}{4}F^{cd}F_{cd}\right)
\\
+2\left[(\mathcal{D}_a \psi) (\mathcal{D}_b \psi)^\dagger+(\mathcal{D}_a \psi)^\dagger(\mathcal{D}_b \psi) -g_{ab} (\mathcal{D}_c \psi)(\mathcal{D}^c \psi)^\dagger-g_{ab} m^2 |\psi|^2  \right]\,,
\end{multline}
\begin{equation}
\nabla_a\left[\left(1+4 \alpha |\psi|^2\right)F^{ab}\right]=i\,q\, \left[(\mathcal{D}^b \psi)\psi^\dagger-(\mathcal{D}^b \psi)^\dagger \psi \right]\,,
\end{equation}
and
\begin{equation}
\mathcal{D}_a \mathcal{D}^a \psi-\alpha F^{cd}F_{cd} \psi-m^2 \psi=0\,.
\label{eq:linearscalar}
\end{equation}
\end{subequations}

In order to understand the static, spherical solutions to the above equations of motion, we use the following standard ansatz
\begin{subequations}
\label{eq:ansatzout}
\begin{equation}
\mathrm{d}s^2=-p(r)\,g(r)^2\,\mathrm{d}t^2+\frac{\mathrm{d}r^2}{p(r)}+r^2 (\mathrm{d}\theta^2 +\sin^2\theta\ \mathrm{d}\phi^2 )
\end{equation}
 For the scalar and Maxwell potential we take
\begin{equation}
A=\Phi(r)\,\mathrm{d}t\,,\qquad \psi=\psi^\dagger=\psi(r)\,.
\end{equation}
\end{subequations}
The equations of motion restricted to our ansatz become
\begin{subequations}\label{EOM}
\begin{align}
&\frac{g}{r^2}\left[\frac{r^2}{g}(1+4\,\alpha\, \psi^2)\Phi^\prime\right]^\prime-\frac{2\,q^2\,\psi^2}{p}\Phi=0\,,
\\
&\frac{1}{r^2g}\left(r^2\,g\,p\,\psi^\prime\right)^\prime+\frac{2\,\alpha\,{\Phi^\prime}^2}{g^2}\psi+\left(\frac{q^2\,\Phi^2}{p\,g^2}-m^2\right)\psi=0\,,
\\
&\frac{g^\prime}{g}-2\,r\,\left(\frac{q^2\Phi^2\psi^2}{p^2g^2}+{\psi^\prime}^2\right)=0\,,
\\
&\frac{1}{r^2g}\left(r\,g\,p\,\right)^\prime-\frac{1}{r^2}+2\,m^2\psi^2+\frac{1+4\,\alpha\,\psi^2}{g^2}{\Phi^\prime}^2=0\,,
\end{align}
\end{subequations}%
where ${}^\prime$ denotes a derivative with respect to $r$. Note that there are second order differential equations for $\Phi$ and $\psi$, but only first order equations for $g$ and $p$.
The event horizon $r=r_+$ is the largest root of $p(r)$, and we will focus on the region outside the horizon, $r\geq r_+$. (The behavior inside the horizon is studied in \cite{dhs}.)
 For numerical convenience we work with a compact radial coordinate
\begin{equation}
z=\frac{r_+}{r}\in(0,1)\,,
\end{equation}
and change variables as
\begin{equation}
p(r)=\left(1-z\right)q_1(z)\,,\quad \Phi(r) = \left(1-z\right)q_2(z)\,,\quad \psi(r)=q_3(z)\quad \text{and}\quad g(r)^2=q_4(z)\,.
\end{equation}
(This imposes the gauge condition that $A_t = 0$ on the horizon.) 
We then solve for $q_1$, $q_2$, $q_3$ and $q_4$ subject to appropriate boundary conditions. At asymptotic infinity, located at $z=0$, we demand 
\begin{equation}
q_1(0)=q_4(0)=1\,,\quad q_3(0)=0\,,\quad\text{and}\quad q_2(0)=\mu
\end{equation}
with $\mu$ being the electrostatic potential.

The hairy black hole solutions depend on several parameters. In addition to the parameters in the action $\{m, q, \alpha\}$, black holes are characterized by their mass $M$ and charge $Q$.
These turn out to be given by
\be
 M = \frac{r_+}{2}[1-\dot{q}_1(0)]\,,
\qquad
Q = r_+\,[\mu-\dot{q}_2(0)]\,,
\ee
where $\dot{}$ denotes a derivative with respect to $z$.
There is a scaling symmetry, so we will present our results using the four dimensionless quantities $\{q/m, \alpha, M\,m,  Q\,m\}$. However, to find the solutions numerically, it is more convenient to use a slightly different set of dimensionless quantities: $\{q/m,\alpha, y_+, \mu\}$, where $y_+ \equiv m \, r_+$ controls directly the area of the black hole event horizon (located at $z=1$ in our compact coordinates).

 At the horizon, smoothness determines the behaviour of all functions, giving a Dirichlet boundary condition, and three Robin boundary conditions for $q_2$, $q_3$ and $q_4$. For concreteness we present the Dirichlet condition which takes the form
\begin{equation}
q_1(1)=1-2 y_+^2 q_3(1)^2-\frac{q_2(1)^2}{q_4(1)} \left[1+4\,\alpha\, q_3(1)^2\right]\,.
\end{equation}
The strategy is now clear: for each value of $\{q/m, \alpha, y_+, \mu\}$ we solve the resulting equations of motion as a boundary value problem with the above boundary conditions. We solve these via a standard relaxation method on a Gauss-Lobatto collocation grid (see \cite{Dias:2015nua} for a review of such numerical methods).

At several points in the main text, we will refer to the entropy and temperature of the black holes. These are given by 
\be
m^2\,S=\pi\,y_+^2\quad\text{and}\quad \frac{T}{m}=\frac{q_1(1)\sqrt{q_4(1)}}{4\pi y_+}\,.
\ee
It is a simple exercise to show that the mass $M$, charge $Q$, chemical potential $\mu$, entropy $S$ and Hawking temperature $T$ obey the first law of black hole mechanics
\begin{equation}
\mathrm{d}M= T\,\mathrm{d}S+\mu\,\mathrm{d}Q\,,
\end{equation}
which we check numerically throughout. All solutions in this manuscript satisfy this relation to at least the $10^{-4}\%$ level of confidence.

Finally, we note that when the scalar field vanishes, \emph{i.e.} $\psi=0$, the only  black hole  is given by the familiar Reissner-Nordstr\"om (RN) solution for which
\begin{equation}
\label{RNsol}
p(r)=p_{\mathrm{RN}}(r)\equiv\frac{(r-r_+)(r-r_-)}{r^2}\,,\quad g(r)=1\,,\quad\text{and}\quad \Phi(r)=\Phi_{\mathrm{RN}}(r)\equiv\left(1-\frac{r_+}{r}\right)\mu
\end{equation}
with $Q=\mu\,r_+$ and $r_{\pm}\equiv M\pm\sqrt{M^2-Q^2}$.
The RN temperature is $T_{\mathrm{RN}}=\frac{r_+-r_-}{4\pi r_+^2}$ and, at extremality, one thus has $r_-=r_+=M=Q$ and $\mu=1$. Note that $r_-/r_+ = \mu^2$.

\subsection{Asymptotic condition}
There is another condition that must be satisfied in order to obtain hairy black holes.
The scalar field will be bound to the black hole only if it  falls off appropriately at infinity. In our gauge with $A_t (r_+) = 0$, and  $A_t(r=\infty) =\mu $, this is only possible if
\be\label{condition}
q^2 \mu^2 \le m^2\,.
\ee
The necessity of this condition can be seen by considering the asymptotic behavior of the scalar field. If $ q^2 \mu^2< m^2$, the scalar field behaves at large radius like
\begin{equation}\label{outpsiinf1}
\psi =\frac{e^{- r \sqrt{m^2-q^2 \mu^2}}}{r^{1+\eta}}\left [b + \mathcal{O}(r^{-1}) \right ],
\end{equation}
for a constant $b$, where 
\begin{equation}\label{outpsiinf2}
\eta \equiv \sqrt{m^2-q^2 \mu^2}\,M-\frac{\mu\,q^2\,(\mu  M-Q)}{\sqrt{m^2-q^2 \mu^2}}\,.
\end{equation}
The exponential decay at large distance is characteristic of a bound state. 

If $ m^2=q^2 \mu^2$, the scalar field still decays exponentially like
\be\label{outpsiinf3}
\psi = \frac{e^{-2 \sqrt{2} \,q \sqrt{\mu }\sqrt{Q-\mu  M}\; \sqrt{r}}}{r^{3/4}}\left[b+\mathcal{O}(r^{-1/2})\right].
\ee
However, if $ q^2 \mu^2 > m^2$, the scalar field oscillates asymptotically indicating that the scalar field is not bound to the black hole. More importantly, such solutions would have infinite energy. 

\section{Linear instability}\label{sec:linear}

The familiar RN metric with $\psi = 0$ is clearly always a solution to our equations of motion \eqref{EOM:S}.  However,  this solution can become unstable to forming scalar hair. This is because  $F^2 < 0$ for an electrically charged black hole, so the last term in the action acts like a negative contribution to the scalar mass. This can become large enough near the horizon to dominate the $m^2$ term in the action.

In this section we determine when this instability sets in using a linearized analysis. In particular, we will take Eq.~(\ref{eq:linearscalar}) and set the metric and gauge field to be those of the RN black hole \eqref{RNsol}. Furthermore, we will take the scalar field $\psi$ to be radially symmetric and Fourier expand in time as
\begin{equation}
\psi(t,r) = \tilde{\psi}(r)\,e^{-i\,\omega\,t}\,,
\end{equation}
which introduces the frequency $\omega$ of the perturbation and brings the scalar equation (\ref{eq:linearscalar}) to the following form
\begin{equation}
\frac{1}{r^2}\left[r^2 p_{\mathrm{RN}}(r)\tilde{\psi}^\prime(r)\right]^\prime+\left\{\frac{\left[\omega+q\,\Phi_{\mathrm{RN}}(r)\right]^2}{p_{\mathrm{RN}}(r)}-m^2+2\,\alpha\,{\Phi_{\mathrm{RN}}^\prime(r)}^2\right\}\tilde{\psi}(r)=0\,.
\label{eq:linear}
\end{equation}

We would like to understand whether finite energy excitations, regular on the future event horizon of the RN black hole, exist for which $\mathrm{Im}\, \omega>0$, in which case we have a mode whose amplitude grows in time and the system develops an instability. Searching for such excitations amounts to studying a generalised eigenvalue problem in $\omega$, which we present in Appendix \ref{sec:Appendix}. Here we present a simple criterion for when RN is unstable, and compute the onset of the instability by looking for $\omega = 0$ modes.
\subsection{The near horizon analysis}\label{sec:linearNH}

Since the RN black hole has a maximum electric field at extremality, we expect that the minimum charge ratio $q/m$ and minimum $\alpha$ needed to herald an instability can be determined by analysing the extremal solution.

The near horizon geometry of the extremal RN black hole takes the direct product form $\mathrm{AdS}_{2}\times S^2$ where $\mathrm{AdS}_{2}$ stands for 2-dimensional anti-de Sitter spacetime. This is best seen by first setting $r_-=r_+$, introducing new coordinates $(\tau,\rho)$ as
\begin{equation}\label{NHcoord}
t = \frac{r_+\,\tau}{\lambda}\,,\quad\text{and}\quad r=r_+(1+\lambda\,\rho)
\end{equation}
and taking the limit $\lambda\to0$. Once we do this, one obtains
\begin{subequations}\label{NHsolution}
\begin{equation}
\mathrm{d}s^2_{\mathrm{AdS}_{2}\times S^2}= L^2_{\mathrm{AdS}_{2}}\left(-\rho^2\mathrm{d}\tau^2+\frac{\mathrm{d}\rho^2}{\rho^2}\right)+r_+^2\,\left(\mathrm{d}\theta^2+\sin^2\theta\,\mathrm{d}\phi^2\right)
\end{equation}
and
\begin{equation}
A_{\mathrm{AdS}_{2}\times S^2}=\mu_{\mathrm{AdS}_2}\,\rho\,\mathrm{d}\tau\,,
\end{equation}
\end{subequations}
where the first factor in the line element corresponds to the two-dimensional AdS$_2$ with $L_{\mathrm{AdS}_{2}}=r_+$ and $\mu_{\mathrm{AdS}_2}=r_+$. The near-horizon solution \eqref{NHsolution} solves \eqref{EOM} with $\psi=0$.

It is a well know fact that \emph{neutral} massive scalar waves propagating on asymptotically AdS spacetimes possess a value for the mass squared below which AdS is unstable and negative energy solutions to the wave equation can be constructed. This is the so-called Breitenl\"ohner-Freedman (BF) bound \cite{Breitenlohner:1982bm,Breitenlohner:1982jf}. In particular, for a neutral massive scalar field in $\mathrm{AdS}_2$ this bound reads
\begin{equation}
m^2_{\mathrm{AdS}_2}L_{\mathrm{AdS}_{2}}^2\geq -\frac{1}{4}\,.
\end{equation}
However, a \emph{charged scalar} field not only gets contributions from bare mass terms in its equation of motion, but also from the gauge fields, since these can act as effective two-dimensional masses. It was  first conjectured in \cite{Denef:2009tp}, and proved in certain cases in \cite{Dias:2010ma}, that the the \emph{full} extreme black hole is unstable with respect to charged perturbations if
\begin{equation}
m^2_{\mathrm{eff}}L_{\mathrm{AdS}_{2}}^2\equiv m^2_{\mathrm{AdS}_2}L_{\mathrm{AdS}_{2}}^2-q^2\mu_{\mathrm{AdS}_2}^2<-\frac{1}{4}\,.
\end{equation}
This is a sufficient, but not necessary condition in general. In the Appendix~\ref{sec:Appendix} we argue that for our case, this condition is also necessary (see, in particular, Sec.~\ref{sec:A3} and the discussion associated to Fig.~\ref{fig:extremalfrequency}). Note that an instability will only be physically acceptable if it is possible to keep $m^2$ positive from the perspective of the asymptotic flat ends, and yet have $m^2_{\mathrm{eff}}L_{\mathrm{AdS}_{2}}^2<-1/4$ in the near horizon $\mathrm{AdS}_{2}\times S^2$ region. 

It remains to compute $m^2_{\mathrm{eff}}L_{\mathrm{AdS}_{2}}^2$ in our particular theory. This is a rather standard procedure and we refer the reader to \cite{Dias:2010ma} for details\footnote{In short, we apply the coordinate transformation \eqref{NHcoord} to the linearized scalar equation \eqref{eq:linear}, set $\omega = \lambda \tilde{\omega}$ and keep only the leading terms in the $\lambda\to 0$ expansion while keeping $\tilde{\omega}$ fixed. Then, one compares the resulting equation to that of a charged, massive scalar living on a rigid AdS$_2$ with mass $m^2_{\mathrm{AdS}_2}$, charge $q$ and frequency $\tilde{\omega}$. From this, we can reconstruct $m^2_{\mathrm{eff}}L_{\mathrm{AdS}_{2}}^2$.}. In our case we find that the $\mathrm{AdS}_2$ BF bound is violated when
\begin{eqnarray} \label{BFviolation}
&& m^2_{\mathrm{eff}}L_{\mathrm{AdS}_{2}}^2+\frac{1}{4}=\frac{1}{4}+(m^2-q^2)L^2_{\mathrm{AdS}_2}-2\alpha<0 \nonumber\\
&& \qquad\qquad\qquad\qquad\qquad \Rightarrow \alpha>\frac{1}{2}\left[\frac{1}{4}+(m^2-q^2)L^2_{\mathrm{AdS}_2}\right]\,.
\label{eq:bound}
\end{eqnarray}
When the background RN black hole is extremal, \emph{i.e.} when $\mu=1$, the bound state condition given in Eq.~(\ref{condition}) simplifies to $m>|q|$, so that the term on the right hand side of the above inequality is always positive. This is essentially the reason why we need the new coupling $\alpha$ if we want to make the RN black hole unstable.

\subsection{The onset of hairy black holes}\label{sec:linearOnset}

When \eqref{BFviolation} is satisfied, the extremal RN black hole is unstable, so the onset of the instability starts at some $Q<M$. This onset can be found by searching for static, finite energy perturbations, so  we set $\omega = 0$ in \eqref{eq:linear}. 

Typically, the onset occurs when $q^2\mu^2 < m^2$. In this case we require that $\psi$ fall off as in \eqref{outpsiinf1} and \eqref{outpsiinf2}.
It is convenient not to work directly with $\psi$, but instead define a new function $\hat{\psi}$ through the relation
\begin{equation}
\psi \equiv  e^{-\sqrt{m^2-q^2\mu^2}\,r}\left(\frac{r_+}{r}\right)^{1+\eta}\hat{\psi}\,.
\label{eq:off}
\end{equation}
Numerically, it is hard to work with infinite domains so we introduce a compact coordinate $y$ given by
\begin{equation}
r=\frac{r_+}{1-y}\,,
\label{eq:ycoord}
\end{equation}
with the horizon located at $y=0$ and asymptotic infinity at $y=1$. The boundary conditions for $\hat{\psi}$ are then found by demanding $\hat{\psi}$ to have a regular Taylor expansion at $y=0$  and $y=1$. This procedure yields rather cumbersome Robin boundary conditions at $y=0$ and $y=1$ which we do not present here.

If we now fix  $\alpha$,  $q/m$, and $m \, r_+$, the equation for $\hat{\psi}$ is a generalized eigenvalue equation in $\mu$. By computing these eigenvalues, we determine a curve in the space of RN black holes that marks the onset of the scalar hair. This is how the blue curve  was generated in Fig.~\ref{fig:phasediag}.

For $q^2 > m^2$, modes with $q^2\mu^2 = m^2$ can also branch off from RN. These are the beginning of the solutions that  we discuss in the next section. To find them, we require that $\psi$ satisfy \eqref{outpsiinf3}
asymptotically,
and set
\begin{equation}
\psi = e^{-2 \sqrt{2}\,q \sqrt{\mu } \sqrt{Q-\mu  M}\; \sqrt{r}}\left(\frac{r_+}{r}\right)^{3/4}\hat{\psi}\,.
\end{equation}
It is again convenient to introduce a compact coordinate
\begin{equation}
r = \frac{r_+}{y^4}\,,
\end{equation}
so that the higher order terms in $r^{-1/4}$ appearing in the expansion \eqref{outpsiinf3} now become integer powers of $y$. The boundary conditions for $\hat{\psi}$ can then be found by assuming that $\hat{\psi}$ has a regular Taylor series at $y=0$ (asymptotic infinity) and $y=1$ (black hole event horizon). They again turn out to be Robin boundary conditions. For fixed $\alpha$ and $q/m$, we regard the equation for $\hat{\psi}$ as an eigenvalue equation for $m \, r_+$, and solve for these eigenvalues. This is how the onset line was generated in Fig.~\ref{fig:qm}. 

In the Appendix~\ref{sec:Appendix} we show that these $\omega=0$ modes indeed mark a transition between stable and unstable perturbations (see, in particular, Sec.~\ref{sec:A2} and the discussions associated to  Figs.~\ref{fig:example}-\ref{fig:3D}).

\section{Maximal warm holes}
 
\begin{figure}[h!]
    \centering
    \vspace{1cm}
    \includegraphics[width=0.85\textwidth]{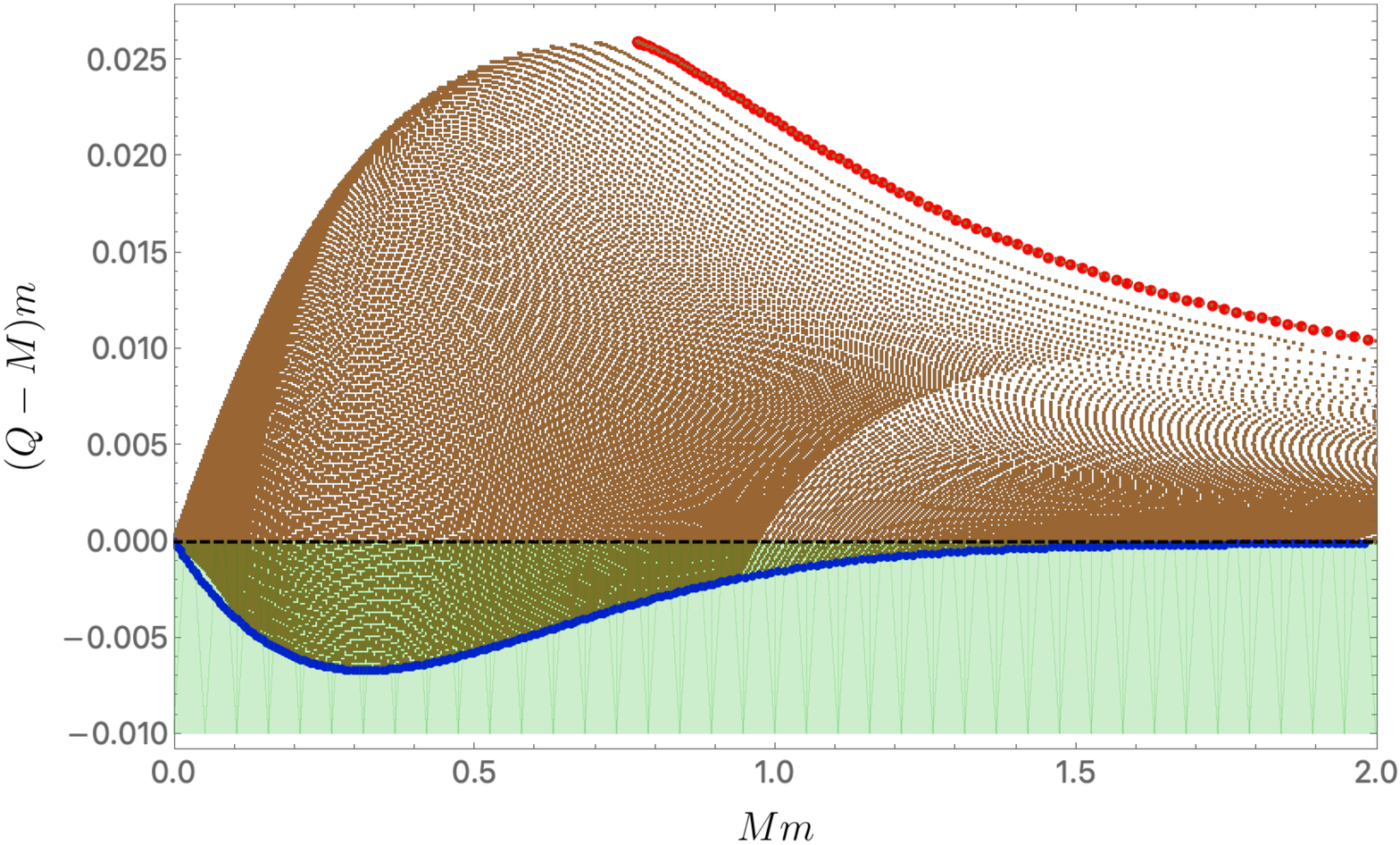}
    \caption{The phase diagram of solutions with $q/m =1$ and $\alpha=1$. Hairy black holes exist in the brown shaded region. The blue line denotes the onset of the scalar instability, and the red line denotes the curve with $\mu =1$. Note that these black holes can slightly exceed the usual extremal bound $Q=M$.  } 
       \label{fig:phasediag}
\end{figure}

We now discuss the full nonlinear solutions, and start with  the case $q/m =1$.\footnote{From now on we assume charges are positive, but our results remain valid if $q$ and $Q$ are replaced by their absolute value.}  So the  condition \eqref{condition} is satisfied for $\mu \le 1$. A phase diagram of these solutions is shown in
 Fig.~\ref{fig:phasediag}, for coupling $\alpha =1$. The green region below the horizontal dashed line with $Q-M=0$ describes the standard RN solutions. The blue line denotes the onset of the scalar instability in RN and thus the merger between the RN and the hairy black holes. The latter exist in the 
 brown shaded region, and the red line denotes the curve $\mu = 1$ which  represents the largest charge  on a black hole of mass  $Mm \gtrsim 0.8$. Notice that the vertical axis is proportional to $Q-M$, so when this is positive, the hairy  black holes exceed the usual extremal limit $Q=M$. It is not surprising that one can create black holes with $Q>M$ by adding matter with $q=m$, since one can also do this with neutral matter. The point is simply that the equation of motion (2.2b) with $q=0$ implies that the conserved charge is $\oint (1+4\alpha|\psi|^2)\star F$. So the electric charge $Q_{\mathcal{H}}$ on the black hole, defined as
\begin{equation}
Q_{\mathcal{H}} \equiv \frac{1}{4\pi} \oint_{\mathcal{B}}\star F\,,
\end{equation}
where $\mathcal{B}$ is the bifurcating Killing surface, will be less than the total charge $Q$ measured at infinity.

\begin{figure}[h!]
    \centering
    \vspace{1cm}
    \includegraphics[width=1\textwidth]{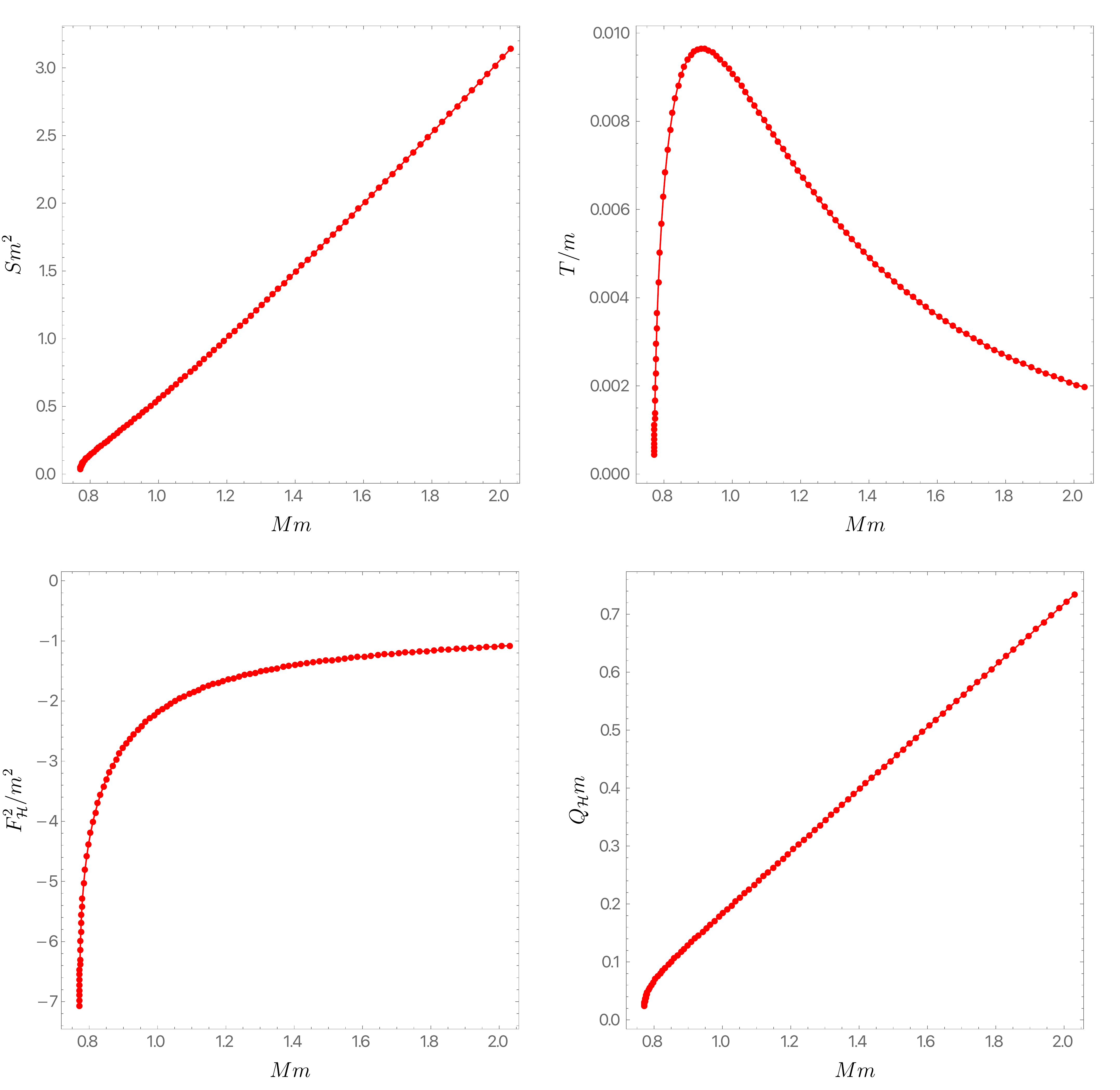}
    \caption{Physical properties of the maximal warm holes along the red line in Fig.~\ref{fig:phasediag}. The plots show the entropy $S$, temperature $T$, $F^2$ on the horizon, and charge on the black hole, $Q_{\mathcal{H}}$, as a function of black hole mass. Note that despite having the maximum charge for a given mass, these black holes have nonzero Hawking temperature! The implications for Hawking evaporation are discussed in section \ref{sec:hawkingeva}.}
       \label{fig:constmu}
\end{figure}

As mentioned in the introduction, the extremal limit of a black hole with scalar hair is often singular, with vanishing horizon area. This is true for the black holes along the left boundary of the phase diagram. However  despite having the largest charge for given mass, the black holes along the red line with $\mu =1 $ are nonsingular ($S\neq 0$),  and  remarkably have nonzero Hawking temperature. This is shown in Fig.~\ref{fig:constmu} which shows various physical properties of the $\mu =1$ black holes including their entropy $S = A/4$, temperature $T$, $F^2$ on the horizon, and charge on the black hole $Q_{\mathcal{H}}$.

The reason these black holes exist can be understood as follows. As one increases their charge (for fixed mass), the region near the horizon behaves as a typical black hole with scalar hair and wants to become singular. However, if the mass is large enough, before one reaches a singular horizon, the asymptotic condition \eqref{condition} is saturated. Since one cannot support scalar hair if this bound is violated, and there are no black holes without hair having $Q>M$, the extremal black hole has $T>0$. This is a new kind of extremal black hole that we are calling a maximal warm hole.

\begin{figure}[h!]
    \centering
    \vspace{1cm}
    \includegraphics[width=0.85\textwidth]{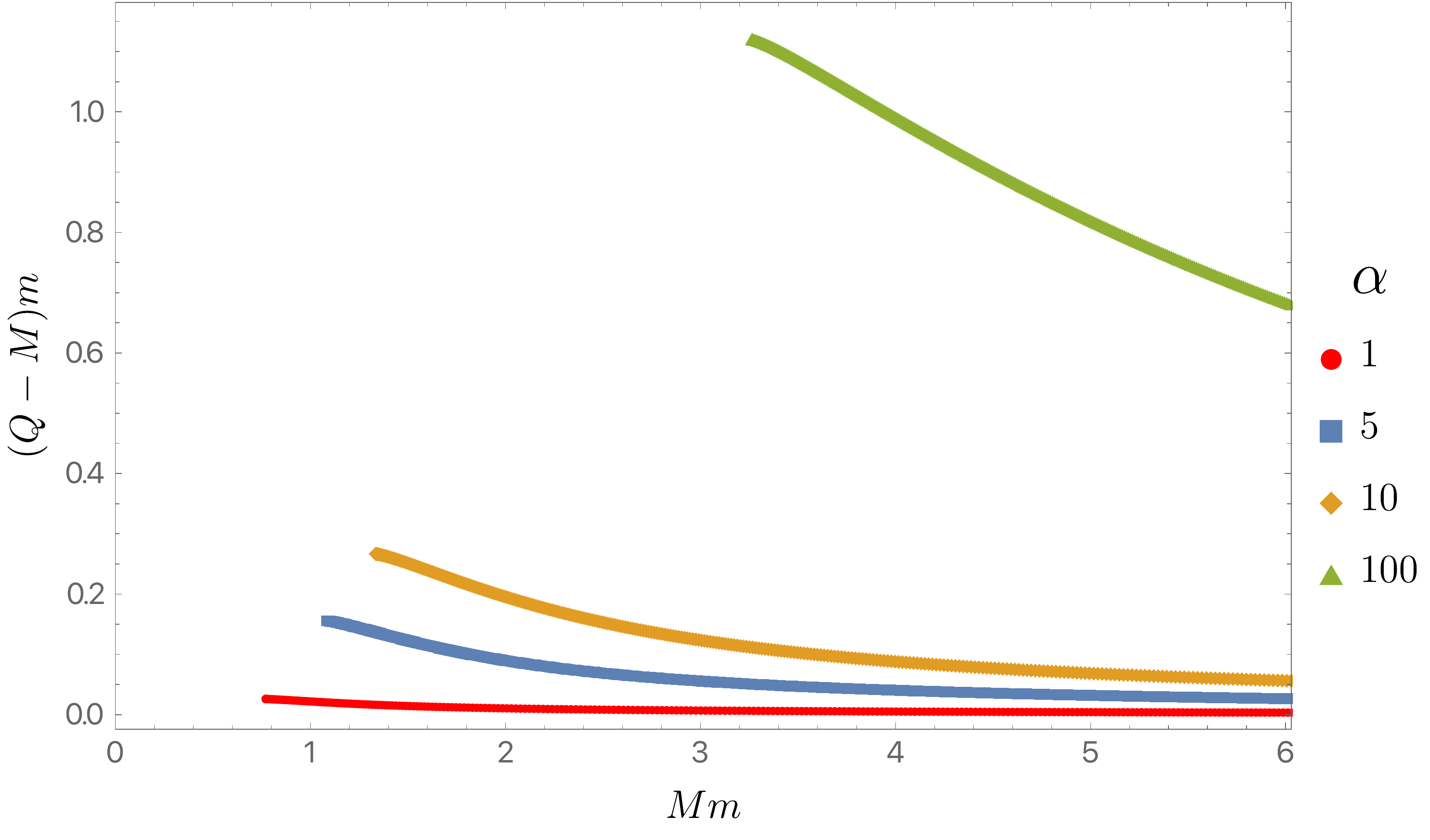}
    \caption{Maximal warm holes in theories with $q=m$ and different couplings $\alpha$. These are all nonsingular ($S> 0$)  black holes with maximum charge and nonzero $T$. As  they approach the solution with minimum mass, $S\to 0$ and $T\to 0$.}
       \label{fig:alpha}
\end{figure}

\begin{figure}[h!]
    \centering
    \vspace{1cm}
    \includegraphics[width=\textwidth]{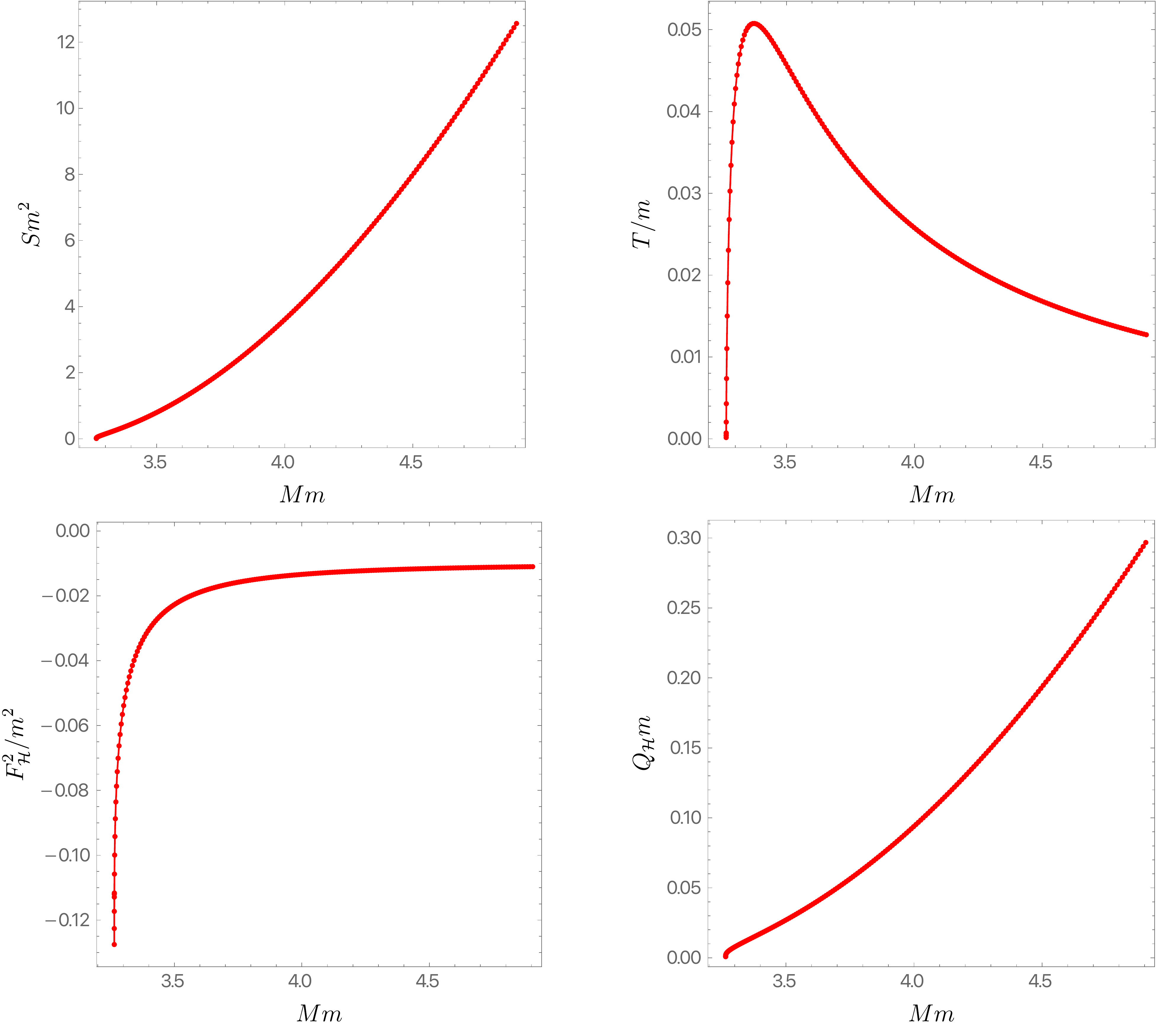}
    \caption{Physical properties of the maximal warm holes with $q=m$ and $\alpha = 100$. Comparing with the $\alpha=1$ case of Fig.~\ref{fig:constmu}, we see that increasing $\alpha$ increases the extremal temperature only slightly (top-right panel), but decreases substantially the fraction of the charge carried by the black hole (bottom-right panel).} 
       \label{fig:tenalpha}
\end{figure}

Increasing the coupling $\alpha$ increases the  charge that these maximal warm holes can carry. But it also increases the minimum mass required for the extremal black hole to be nonsingular. Both of these effects are shown in Fig.~\ref{fig:alpha} which shows the maximal warm holes in theories with $q=m$ and different couplings $\alpha$. These curves all have $\mu = 1$ and generalize the red curve in Fig.~\ref{fig:phasediag} to larger $\alpha$. The physical properties of these black holes are qualitatively similar to Fig.~\ref{fig:constmu}. In particular,  they are all nonsingular with nonzero Hawking temperature. For example, the properties of the black holes when $\alpha = 100$ are shown in Fig.~\ref{fig:tenalpha}. Notice that increasing $\alpha$ increases the extremal temperature only slightly (top-right panel), but greatly decreases the fraction of the charge $Q_{\mathcal{H}}$ that is carried by the black hole (bottom-right panel). Most of the charge is now in the scalar hair, which is not surprising since we have increased the scalar instability.

\begin{figure}[h!]
    \centering
    \vspace{1cm}
    \includegraphics[width=1\textwidth]{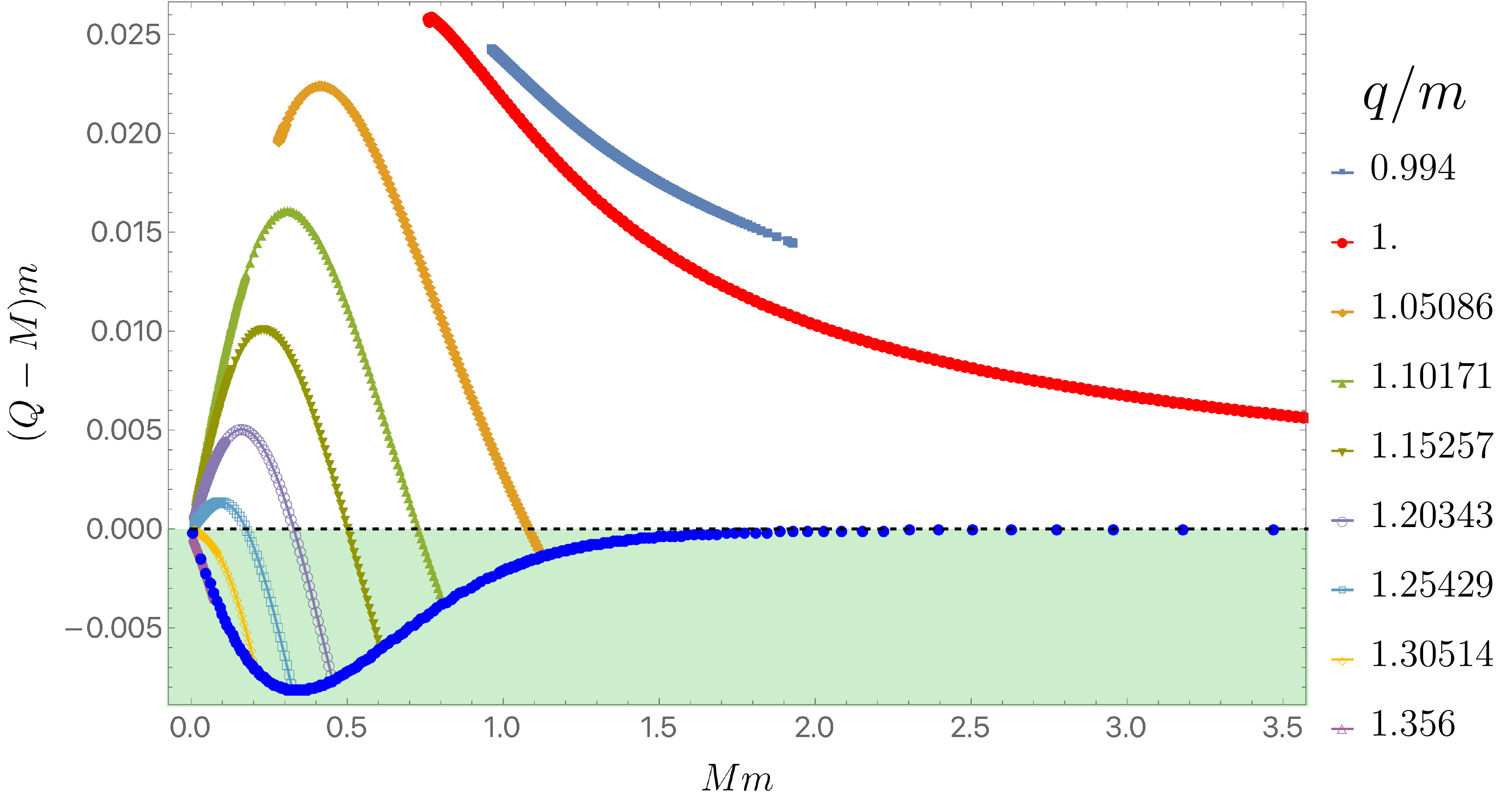}
    \caption{Black holes with $q\mu = m$ as a function of $q/m$, with $\alpha=1$. When $Q>M$, these are maximal warm holes. The green shaded region denotes RN black holes, and the bottom blue curve denotes the onset of their instability when $q\mu = m$. For masses outside the range of the  maximal warm holes, the extremal black hole is singular.} 
       \label{fig:qm}
\end{figure}

Next we return to $\alpha = 1$, and consider the effects of changing $q/m$. The existence of maximal warm holes  turns out to be very sensitive to this parameter. 
The smooth black holes with maximum charge for given mass again have the maximum possible potential difference $\mu$ allowed by \eqref{condition}. They are shown in Fig.~\ref{fig:qm}, and  all have $T > 0$ (except the leftmost point that approaches $S\to 0$ and $T\to 0$).

This figure has several interesting features. First,  black holes with $q/m > 1$ scalar hair only exist when the black hole is small enough. This can be understood as follows. If we increase $q/m > 1$, the maximum value of $\mu$ is reduced to  $\mu \le m/q < 1$. Since the maximum allowed $\mu$ is reduced, the maximum electric field on the horizon is also reduced. But for the RN black hole to become unstable, we need a large enough electric field. Since the electric field increases as one decreases the size of the black hole, only small black holes can have this kind of hair. Second, the mass where maximal warm holes become singular rapidly decreases to zero as $q/m$ increases, and for $q/m \gtrsim 1.1$, maximal warm holes can have arbitrarily small mass. This is also easy to understand: increasing $q/m$ decreases the maximum allowed $\mu$, so this maximum is reached sooner, before the horizon becomes singular. Third, the maximum charge the hairy black hole can carry also decreases as $q/m$ increases, and for $q/m \gtrsim 1.3$, it falls below $Q = M$. At this point, the maximum charge black hole is the usual RN solution with no scalar hair.
However, when they exist, the hairy black holes always have larger entropy than a RN solution with the same $M$ and $Q$. As one increases $Q$ for fixed $M$, the RN solution becomes unstable as before, but if one continues to increase $Q$, one reaches a point where the hair no longer exists and the solution returns to RN.

Next we consider decreasing $q / m < 1$. This increases the maximum allowed $\mu$, making it easier for the horizon to become singular before reaching this limit. So the minimum mass required for a maximal warm hole increases, as shown in Fig.~\ref{fig:qm} for the case $q/m=0.994$. There is also  a maximum mass, but unlike the case $q/m > 1$, it is not because they no longer satisfy $Q > M$. Instead, it is because a solution with $m = q\mu$  requires $Q > \mu M$; see \eqref{outpsiinf3}. This constraint was not an issue when $\mu \le 1$, but since we have increased $\mu$, this constraint is violated for large $M$ and the extremal limit again becomes singular. The finite range of masses for which the black hole has a smooth extremal limit rapidly shrinks as we decrease $q/m$ and vanishes completely for $q/m \lesssim 0.99$.
 When the maximal warm holes only exist for large enough masses (as in the top three curves of Fig.~\ref{fig:qm}), the singular extremal black holes lie along curves that extend from the maximal warm hole with smallest mass to $Q=M=0$. For $q/m<1$, they also extend from the maximal warm hole with largest mass to arbitrarily large $M$.

\begin{figure}[h!]
    \centering
    \vspace{1cm}
    \includegraphics[width=1\textwidth]{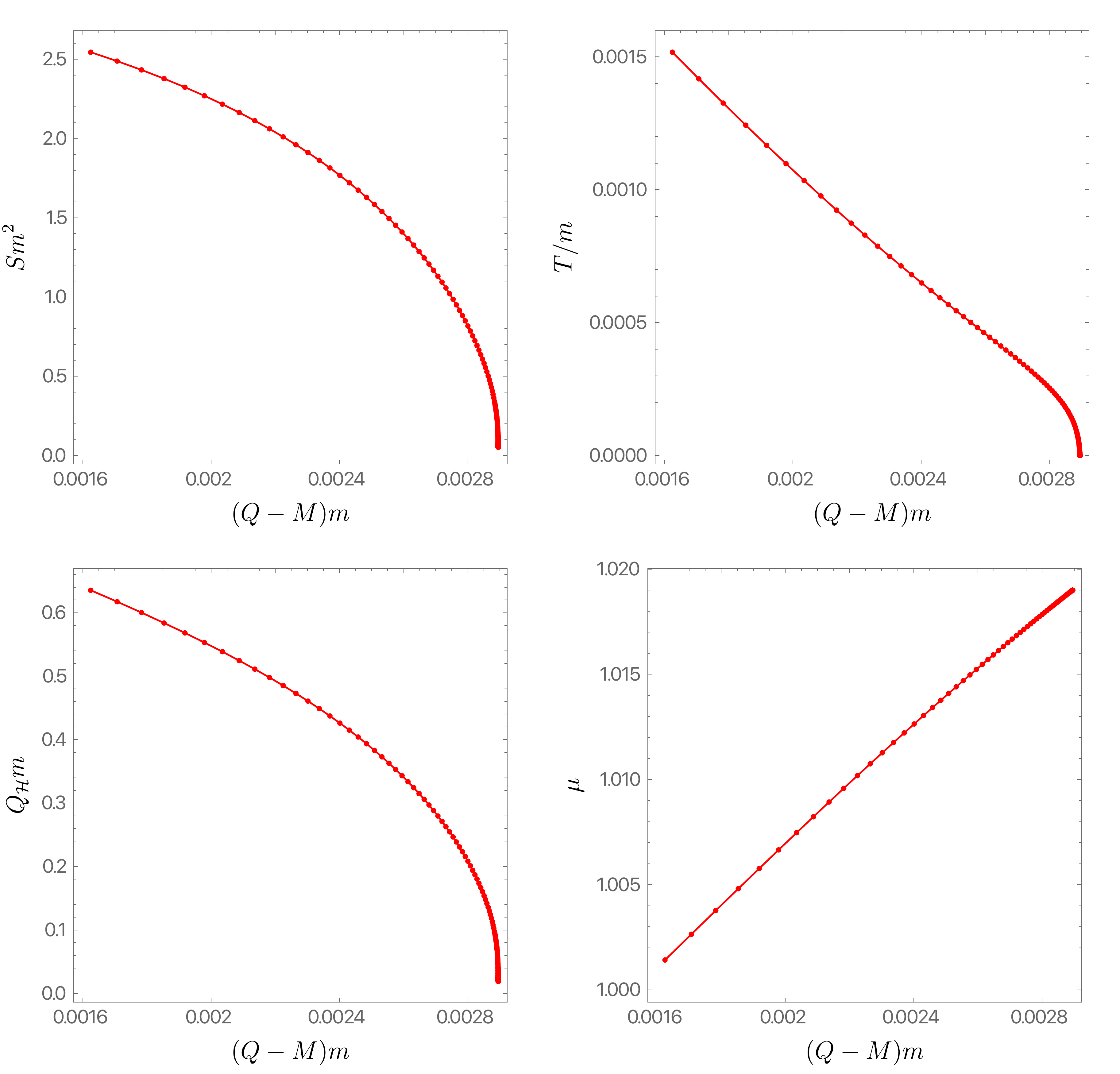}
\caption{Physical properties of black holes approaching extremality, with $q = m/2$, $\alpha = 1$, and $Mm = 1$. This is a representative example of $q/m \lesssim 0.99$ solutions where the maximum charge hairy black holes always approach a singular extremal solution.}
       \label{fig:halfq2}
\end{figure}

For smaller scalar field charges, \emph{i.e.} $q/m \lesssim 0.99$, there are no  nonsingular extremal black holes. In a phase diagram like Fig.~\ref{fig:qm}, hairy black holes with this scalar charge are bounded from above by a single curve that describes singular extremal black holes that extends from $Q=M=0$ to arbitrarily large $Mm$. We might then ask what happens \emph{e.g.} to a nonextremal hairy black hole family with fixed $Mm$ as it approaches the singular extremal curve. The evolution of the physical properties of such a black hole with $q/m = 1/2$ and $Mm =1$ as it approaches extremality are shown in Fig.~\ref{fig:halfq2}. (Other choices of mass $Mm$ and small $q/m$ are similar.) One sees that both the black hole entropy and temperature go to zero in the extremal limit (largest $(Q-M)m$ solution).  The charge on the black hole also vanishes in this limit, since any residual charge would produce a diverging Maxwell field increasing the scalar instability. Note that even though the  condition \eqref{condition} allows $\mu \le 2$ in this case, the solution becomes singular when $\mu$ is only slightly larger than one.

As illustrated in Fig.~\ref{fig:qm}, the only case which allows maximal warm holes to have arbitrarily large mass is the original one we studied with $q=m$ (see Fig.~\ref{fig:phasediag} for $\alpha=1$). The reason for this is that, from \eqref{condition}, the maximum allowed $\mu$ is then  $\mu = 1$  which is just the potential for an extremal RN black hole of any mass. This has two consequences. First, since $q=m$ is  on the threshold of charged superradiance for extremal RN, any extra source of instability (such as the scalar-Maxwell coupling we added) will cause the scalar field to condense (see also Eq.~(\ref{eq:bound})). One does not need the electric field to be ``large enough" in this case. Second, once the black hole has $Q> M$, the second constraint ($Q > \mu M$) that follows from \eqref{outpsiinf3} is satisfied for all $M$. So there is no upper limit on the mass. 

\subsection{Hawking evaporation\label{sec:hawkingeva}}

Typically, if a theory does not have particles with $q > m$, a near extremal black hole will Hawking radiate neutral massless particles such as gravitons and become extremal. Since an extremal RN black hole has zero Hawking temperature, it is a stable endpoint for this process. For some dilatonic black holes with singular extremal limits, the Hawking temperature does not go to zero at extremality \cite{Garfinkle:1990qj}. But in those cases, it has been shown that evaporation stops because the effective potential in the scalar wave equation does not vanish on the horizon as usual in the extremal limit \cite{Holzhey:1991bx}. Since the horizon is at $r_\star = -\infty$ in the usual ``tortoise" coordinate in which the wave equation is simple, this produces an infinite potential barrier allowing no particles to escape.

Since maximal warm holes are smooth black holes with maximal charge and nonzero temperature, we need to find another scenario for the endpoint of their Hawking evaporation.
We will not perform a complete analysis  including the potentials outside the horizon. Instead, we give a simple plausible explanation for why these black holes will \emph{not} form naked singularities, despite the fact that they have maximal charge and nonzero temperature.

Consider first the case $\alpha =1$ and $q=m$. Since the temperature of the hairy black holes is low, charged particles are only created by the Schwinger mechanism with a rate proportional to $e^{-\pi m^2/qE}$, while neutral photons and gravitons are  produced thermally. Since the photons acquire a mass inside the charged condensate, they will be surpressed compared to gravitons. Nevertheless, since charged particle emission appears exponentially suppressed, one might expect that in the late stages of Hawking evaporation, the black hole will lose mass but not charge. Comparing the scales on the horizontal and vertical axes in  Fig.~\ref{fig:phasediag} this would correspond to an essentially vertical line in the figure. So if $Mm$ is large enough, Hawking evaporation would appear to end on the red line. But since these black holes have nonzero temperature, they would appear to keep radiating. This is the puzzle we want to resolve.

\begin{figure}[h!]
    \centering
    \vspace{1cm}
    \includegraphics[width=1\textwidth]{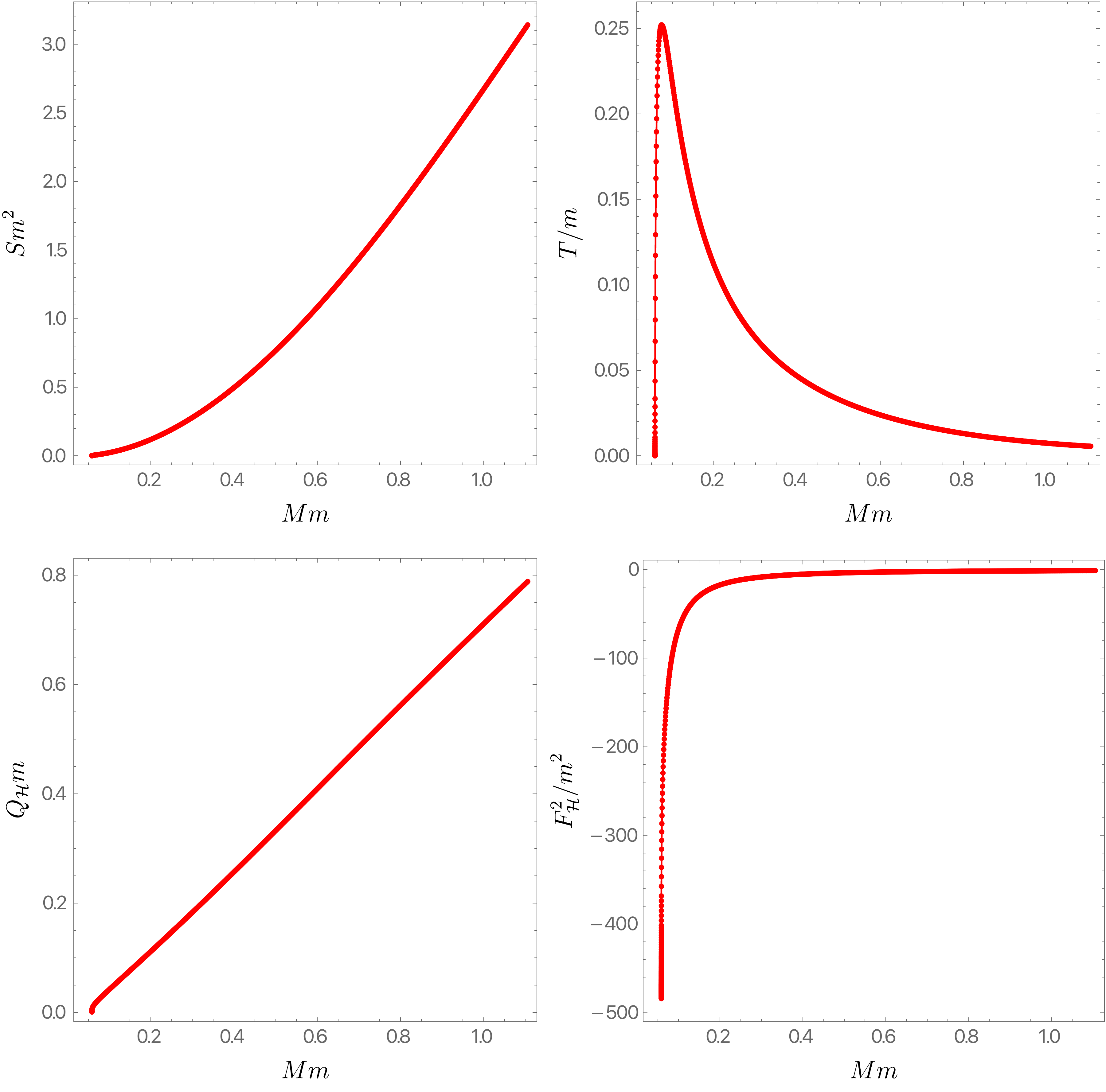}
    \caption{Physical properties of the hairy black hole with $q=m$ and $\alpha = 1$ are shown along a line of constant $(Q-M)m = 5\times 10^{-3}$. In the late stages of Hawking evaporation, the black hole is expected to approximately follow such a line with decreasing $M$.} 
       \label{fig:evap}
\end{figure}

The resolution is that the rate of charged particle production is not actually exponentially suppressed, since all the factors in the Schwinger exponent are order one: we have assumed $q = m$, and Fig.~\ref{fig:constmu} shows that $E/m \sim O(1)$. In contrast, the temperature is $T \sim 10^{-3}$ so the rate of thermal radiation would be proportional to $T^4 \sim 10^{-12}$ and is highly suppressed. Thus the late stages of Hawking radiation will be dominated by the production of $q = m $ particles which should keep $Q-M$ approximately constant. As a result, Hawking radiation causes the black hole to evolve along a horizontal line in Fig.~\ref{fig:phasediag}, rather than a vertical line. This ends in a singular solution as expected.  The physical quantities evolve as shown in Fig.~\ref{fig:evap}. Note that the charge on the black hole goes to zero linearly with the mass, as expected from the production of $q=m$ particles.

Now suppose $q \ne m$. If we increase $q/m$ above 1.1,  we have seen (see discussion of Fig.~\ref{fig:qm}) that there are no singular extremal black holes. But 
Hawking radiation of these hairy black holes is likely to again be dominated by charged particle emission which  will decrease the black hole charge more than its mass. So the black hole will evolve away from extremality. 
On the other hand, if we decrease $q/m$ below $.99$ even charged particle emission will increase $Q-M$, so evaporating black holes will always follow an essentially vertical line in a phase diagram like Fig.~\ref{fig:phasediag}. But we have seen (Fig.~\ref{fig:qm}) that in this case the maximal warm holes disappear  and all extremal limits are singular. 

Thus when $\alpha \approx 1$, the natural endpoint of Hawking evaporation is either a singular extremal solution or possibly a neutral black hole that evaporates completely. The physics of the singular endpoint will of course require a complete quantum theory of gravity. However, the story changes when we increase $\alpha$, since this decreases the electric field on the horizon and increases the black hole temperature. Eventually (certainly before $\alpha = 100$) the electric field becomes too small to create charged particles, and Hawking radiation is dominated by thermal gravitons. Thus we are again faced with the question of what happens when these black holes evaporate past extremality.

Since the black hole is evaporating but not loosing charge, the horizon area will shrink and the potential difference $\mu$ between the horizon and infinity should increase. But $\mu$ was already at the maximum value that allows static scalar hair. So the evaporation past extremality will cause the scalar hair to become unbound and start radiating to infinity. At this point there are a couple possible outcomes depending on how much scalar field is radiated away. If the scalar field only radiates enough to recover $\mu = 1$, the evolution will essentially follow the $\mu = 1$ curve as $M$ decreases. At the other extreme, all the hair could classically radiate away leaving a RN black hole. (This option is only possible if the resulting black hole is classically stable.) Finally, it is possible that a fraction of the hair is radiated leaving a hairy black hole with $\mu < 1$. We will leave it to future investigations to determine which of these possibilities the black hole actually follows. But notice that in no case does the black hole immediately turn into a naked singularity. 

\section{Solitons}

Unlike analogous theories with neutral scalars \cite{Herdeiro:2019oqp},
the theory we are considering also admits soliton solutions, \emph{i.e.}  regular horizonless solutions. For completeness we describe them in this section. We will see that their mass and charge satisfy $Q^2 < M^2$ so they coexist with RN black holes. But unlike other systems with scalar condensation, these solitons are not the zero horizon radius limit of the hairy black holes studied in the previous section. 

Since solitons have no horizon, we have to change our ansatz (\ref{eq:ansatzout}) which was tailored to enforce a zero of $p(r)$. In this section, we thus consider the gravitational ansatz
\begin{equation}
\mathrm{d}s^2=-f(r)\,\mathrm{d}t^2+\frac{\mathrm{d}r^2}{g(r)}+r^2(\mathrm{d}\theta^2 +\sin^2\theta\ \mathrm{d}\phi^2 )\,,
\end{equation}
for the soliton with $r\in(0,+\infty)$. As before, for the Maxwell and scalar fields we take
\begin{equation}
A=\Phi(r)\,\mathrm{d}t \quad \text{and}\quad  \psi = \psi^\dagger = \psi(r)\quad \,.
\end{equation}
The equations of motion read
\begin{subequations}
\begin{align}
& \frac{1}{r^2}\sqrt{\frac{g}{f}}\left(\sqrt{f\,g}\,r^2\psi^\prime\right)^\prime+\left(\frac{q^2\,\Phi^2+2\,\alpha\,g\,{\Phi^\prime}^2}{f}-m^2\right)\psi=0\,,\label{eq:firstsoliton}
\\
& \frac{1}{r^2}\sqrt{\frac{g}{f}}\left[\sqrt{\frac{g}{f}}\left(1+4\,\alpha\,\psi^2\right)r^2\Phi^\prime\right]^\prime-\frac{2\,q^2\,\psi^2}{f}\Phi=0\,,\label{eq:secondsoliton}
\\
& \frac{1}{r^2}\left(r\,g\right)^\prime-\frac{1}{r^2}+\frac{g}{f}(1+4\,\alpha\,\psi^2){\Phi^\prime}^2+2\,q^2\,\psi^2\frac{\Phi^2}{f}+2m^2\psi^2+2\,g\,{\psi^\prime}^2=0\,,\label{eq:thirdsoliton}
\\
&\frac{g}{r^2\,f}\left(r\,f\right)^\prime-\frac{1}{r^2}+\frac{g}{f}(1+4\,\alpha\,\psi^2){\Phi^\prime}^2-2\,q^2\,\psi^2\frac{\Phi^2}{f}+2m^2\psi^2-2\,g\,{\psi^\prime}^2=0\,.\label{eq:lastsoliton}
\end{align}
\end{subequations}
We can now use \eqref{eq:lastsoliton} to express $g$ as a function of $f$, $\psi$, $\Phi$ and their first derivatives:
\begin{equation}
g = \frac{2 r^2 \psi^2\left(q^2\Phi^2-m^2f\right)+f}{\left(r\,f\right)^\prime+(1+4\,\alpha\,\psi^2)r^2{\Phi^\prime}^2-2 r^2\,f\,{\psi^\prime}^2}\,.
\end{equation}
This expression for $g$ can now be plugged in (\ref{eq:firstsoliton})-(\ref{eq:lastsoliton}) to reduce the problem to studying a system of three second order coupled ordinary differential equations for $f$, $\Phi$ and $\psi$.

At the spacetime origin, located at $r=0$, we impose regularity, which amounts to requiring
\begin{equation}
f^\prime(0)=\psi^\prime(0)=\Phi^\prime(0)=0\,.
\end{equation}
(Note in particular that these conditions imply  $g(0)-1=g^\prime(0)=0$, as required.)
At the asymptotic boundary we demand
\begin{equation}
\lim_{r\to+\infty}\psi = 0\,,\quad\lim_{r\to+\infty}f=1\quad \text{and}\quad \lim_{r\to+\infty}\Phi=\mu\,.
\end{equation}

We now introduce a compact radial coordinate $y$ defined as
\begin{equation}
y=\frac{m\,r}{1+m\,r}
\end{equation}
so that $y\in(0,1)$ with $y=0$ being the regular center and $y=1$ the spatial infinity.

The moduli space of solutions is then three-dimensional depending on $\{\mu,\alpha,q/m\}$ or alternatively $\{m M,\alpha,q/m\}$. However, as we shall shortly see, these parameters do not uniquely parametrize a soliton. Therefore, we will use instead the value of the scalar field at the origin, $\psi_0\equiv\psi(0)$, to move along the moduli space and determine $\{m M,m\,Q\}$ at fixed $\{\alpha,q/m\}$. It turns out that $\psi_0$ is one-to-one with the soliton solutions, at fixed $\{\alpha,q/m\}$.

\begin{figure}[h!]
    \centering
    \vspace{1cm}
    \includegraphics[width=1.0\textwidth]{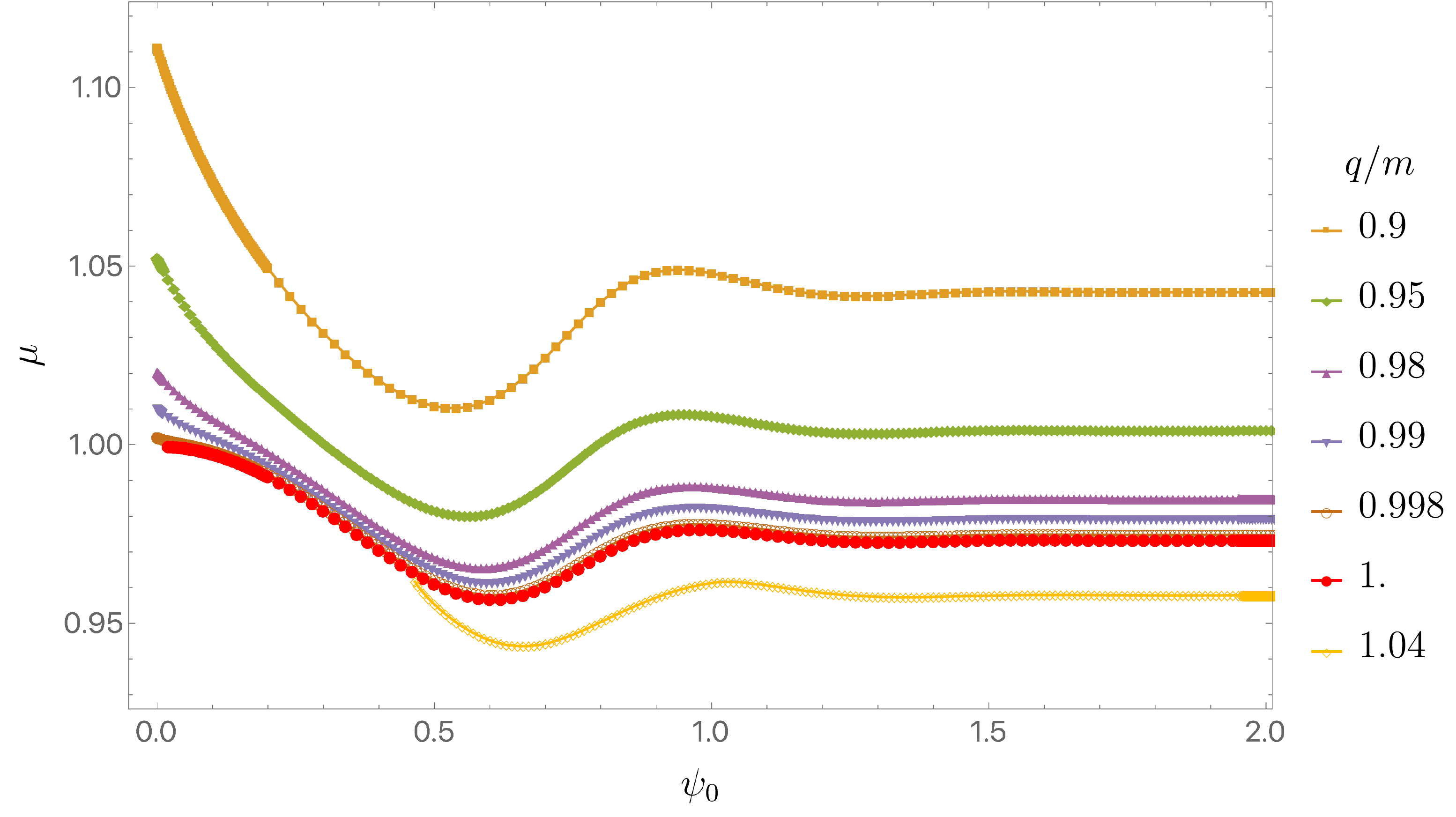}
    \caption{Chemical potential $\mu$ of the solitons as a function of $\psi_0$ at fixed $\alpha=1$. The legend shows curves with different values of $q/m$.} 
       \label{fig:solitons}
\end{figure}

In Fig.~\ref{fig:solitons} we plot the chemical potential $\mu$ as a function of $\psi_0$ for fixed $\alpha=1$ and for several values of $q/m$. The behaviour at large $\psi_0$ is consistent with the following functional form
\begin{equation}
\mu = \mu_{\infty}+\hat{\mu}_{\infty}e^{-\alpha\,\psi_0^2}\sin\left(\Omega_{\infty}\,\psi_0^2+\gamma_{\infty}\right)\,.
\end{equation}
For instance, for $q/m=1$ we find $\mu_{\infty}\approx0.9736$, $\hat{\mu}_{\infty}\approx0.0274$, $\Omega_{\infty}\approx 4.061$ and $\gamma_{\infty}\approx-2.70745$. The above asymptotic expression was inspired by the work developed in \cite{Bhattacharyya:2010yg}, where a class of supersymmetric solitonic solutions was studied in great detail.

\begin{figure}[h!]
    \centering
    \vspace{1cm}
    \includegraphics[width=1.0\textwidth]{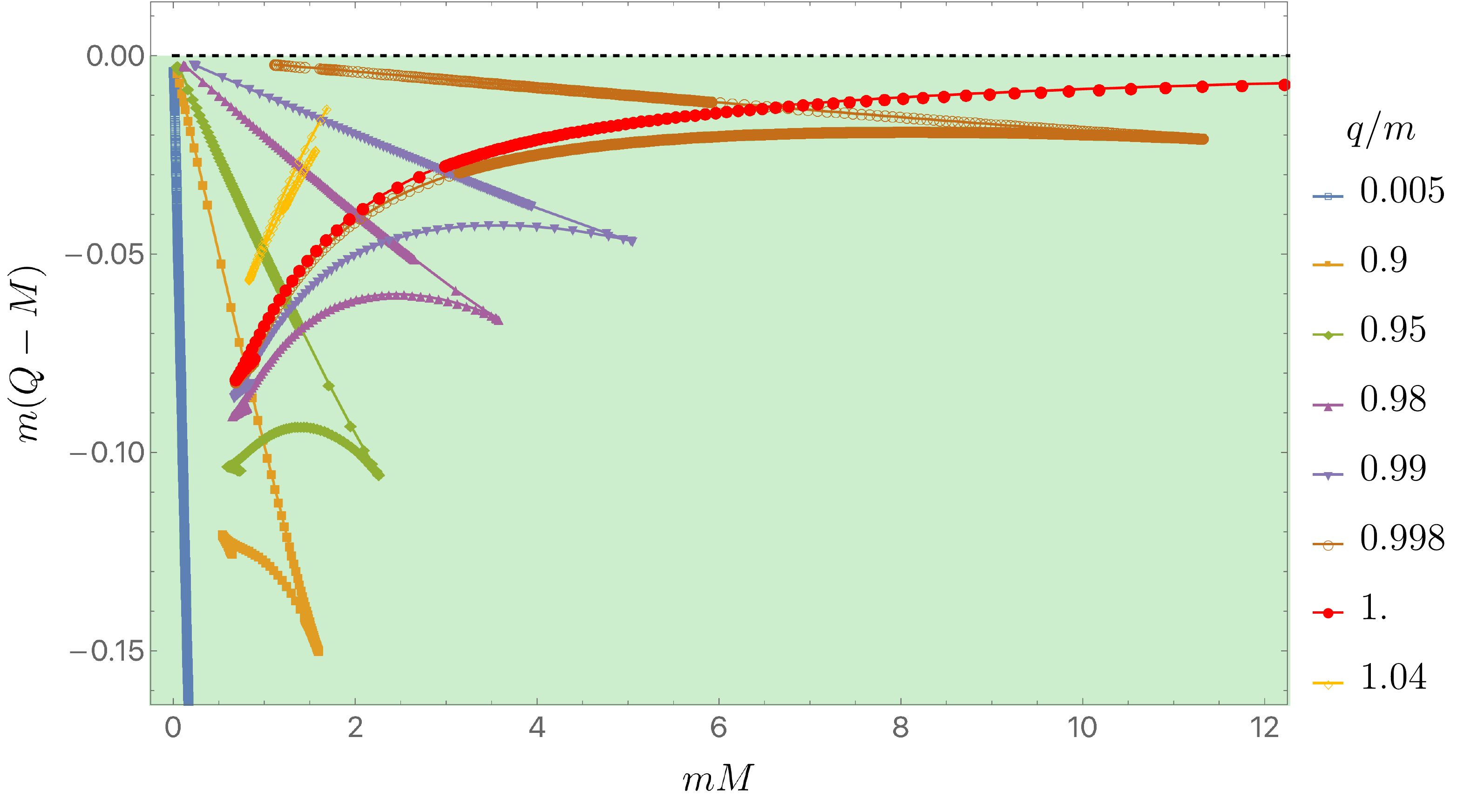}
    \caption{Moduli space of solitonic solutions for fixed $\alpha=1$, and for several values of $q/m$ labelled on the left. The green region indicates where RN black holes exist. } 
       \label{fig:moduli_solitons}
\end{figure}

Each of the oscillations in Fig.~\ref{fig:solitons} is mapped into characteristic swallowtail curves in the corresponding phase diagram of Fig.~\ref{fig:moduli_solitons}. This Fig.~\ref{fig:moduli_solitons} has $\alpha=1$ and serves to illustrate that the properties of solitonic solutions in this theory are somehow intricate. For any value of $0<|q|/m<1$ we find that solitons only exist in a window of masses $ M\in(M_{\min},M_{\max})$. 
For each $q/m$ curve, $M_{\min}$ in Fig.~\ref{fig:moduli_solitons} corresponds to approach $\psi_0\to 0$ in Fig.~\ref{fig:solitons}.
As we decrease $|q|/m$ towards zero, $M_{\min}$ appears to approach $0$ and the curve becomes increasingly steep (see for instance the curve with $|q|/m=5\times 10^{-3}$ in Fig.~\ref{fig:moduli_solitons}). On the other hand, for each $q/m$, $M_{\max}$ in Fig.~\ref{fig:moduli_solitons} corresponds to the first minimum in the corresponding curve of Fig.~\ref{fig:solitons}.  

The solution with $q=m$ is special. In this case when we let $\psi_0\to0$ we approach $M\to+\infty$ and the line $Q-M\to 0$ from below. But there is still a minimum value of the mass $M_{\min}$, which is given by the corresponding minimum in Fig.~\ref{fig:solitons}. 

Finally, we also found solitons with $|q|/m>1$. In this case, solitons again exist in a window $M\in(M_{\min}, M_{\max})$, with the window shrinking as we increase $|q|/m$ and disappearing altogether at a critical value of $|q|=q_c$. For $\alpha=1$, we find that $q_c\simeq 1.05\,m$. For $|q|>m$, we also find that $\psi_0$ never approaches zero, and is instead cut off by a value $\psi_0^c$ at which point the solution becomes singular since the Kretschmann curvature scalar at the origin grows unbounded.

By comparing the mass and charge of the solitons in Fig.~\ref{fig:moduli_solitons} with the mass and charge of the hairy black holes, one finds that they do not overlap. So the solitons cannot be viewed as the limit of a hairy black holes as $r_+\rightarrow 0$.
\section {Discussion}

We have shown that just by adding a simple coupling between a charged scalar field and a Maxwell field, one can change some basic properties of four-dimensional, asymptotically flat, extremal black holes. In particular, for a range of parameters the black hole with maximum charge (for given mass) has a smooth horizon with nonzero Hawking temperature. We have called these objects   maximal warm holes.

The existence of maximal warm holes raises a number of questions. We have (partially) addressed perhaps the most obvious one concerning the endpoint of Hawking evaporation. But in addition to gaining a more complete understanding of this process, there are a number of other questions which we leave for future investigation. These include the following:

\begin{enumerate}

\item  What characterizes the class of theories in which maximal warm holes occur?

\item Do maximal warm holes have implications for  black hole physics besides Hawking radiation?

\item Can maximal warm holes exist in more than four spacetime dimensions?

 \item Are there asymptotically anti-de Sitter examples of maximal warm holes? If so, what are the implications for the AdS/CFT correspondence? They do not exist in the simplest models of holographic superconductors \cite{Horowitz:2009ij}, but they might exist in theories with additional interactions. 
 
 \item Are there asymptotically de Sitter examples of maximal warm holes? This seems unlikely since one would need to ensure that there is no flux across both the cosmological and event horizons. 
 
 \item How does the addition of rotation affect maximal warm holes? It is known that Kerr black holes can develop massive scalar hair near extremality even without additional interactions \cite{Herdeiro:2014goa,Herdeiro:2015gia,Chodosh:2015nma,Chodosh:2015oma}. Can extremal neutral black holes have nonzero temperature?
 
\end{enumerate}

\noindent We hope to report on some of these questions in the future.

\subsection*{Acknowledgments}
 O.J.C.D. acknowledges financial support from the STFC  Grants ST/P000711/1 and ST/T000775/1. 
The work of G.~H. was supported in part by NSF Grant PHY-2107939. J.~E.~S has been partially supported by STFC consolidated grants ST/P000681/1, ST/T000694/1.
The numerical component of this study was partially carried out using the computational facilities of the Fawcett High Performance Computing system at the Faculty of Mathematics, University of Cambridge, funded by STFC consolidated grants ST/P000681/1, ST/T000694/1 and ST/P000673/1. The authors further acknowledge the use of the IRIDIS High Performance Computing Facility, and associated support services at the University of Southampton, in the completion of this work.

\appendix

\section{Unstable modes}\label{sec:Appendix}

In this Appendix we study general linearized scalar perturbations to RN in our theory. As described in Sec.~\ref{sec:linear}, these frequency dependent linearized modes satisfy \eqref{eq:linear}. We will confirm that the static modes discussed in Sec.~\ref{sec:linearOnset} indeed mark the transition between stable and unstable perturbations, and compute the growth rates of the unstable modes. We also argue that RN is unstable if and only if the near horizon condition \eqref{eq:bound} is satisfied.

\subsection{Near horizon geometry modes}\label{sec:A1}

We begin by finding modes in the near horizon geometry \eqref{NHsolution}. In Sec.~\ref{sec:linearNH} we found that the violation of the near horizon BF bound in the extremal RN geometry allows us to predict when an instability occurs (\emph{i.e.} when $\mathrm{Im}\,\omega\geq 0$). In this subsection we complement this analysis: we perform a near horizon analysis of the linear scalar equation (\ref{eq:linear}) to find that when the near horizon BF is preserved the perturbations close to extremality have a quasinormal mode spectrum whose imaginary part {\it vanishes} at extremality. 

In order to do this, we introduce the off-extremality parameter
\begin{equation}
\sigma\equiv 1-\frac{r_-}{r_+}=1-\mu^2\,,
\end{equation}
and set
\begin{equation}
\omega = \tilde{\omega}\,\sigma\,,\quad \text{and}\quad z=\frac{r-r_+}{\sigma\,r_+}\,.
\end{equation}
We wish now to take the limit $\sigma\to0$ in~\eqref{eq:linear}, while keeping $z$ and $\tilde{\omega}$ fixed. This amounts to zooming in near $r=r_+$. The resulting equation reads
\begin{equation}
z(1+z)\frac{\mathrm{d}^2\tilde{\psi}}{\mathrm{d}z^2}+(1+2\,z)\frac{\mathrm{d}\tilde{\psi}}{\mathrm{d}z}+\left[\frac{r_+^2(q\,z+\tilde{\omega})^2}{z(1+z)}+2\alpha-m^2r_+^2\right]\tilde{\psi}=0\,.
\end{equation}
The above equation can be related to Euler's hypergeometric differential equation, whose solution can be expressed as a sum of two Gauss hypergeometric functions. We are interested in the solution that is regular across the future event horizon, which takes the simple form
\begin{subequations}\label{appNH}
\begin{equation}
\tilde{\psi}(z) = (1+z)^{i\,r_+(q-\tilde{\omega})}z^{-i\,r_+\,\tilde{\omega}}{}_2F_1\left(a_-;a_+;c;-z\right)
\end{equation}
where ${}_2F_1(a;b;c,z)$ is the Gauss hypergeometric function and
\begin{align}
a_{\pm}& \equiv \frac{1}{2}+i\,r_+\,(q-2\tilde{\omega})\pm\sqrt{m^2_{\mathrm{eff}}L_{\mathrm{AdS}_{2}}^2+\frac{1}{4}}\,,
\\
c& \equiv 1-2\,i\,r_+\,\tilde{\omega}\,.
\end{align}
\end{subequations}

We can now expand \eqref{appNH} at large $z$ to find
\begin{equation}\label{appNHlargez}
\tilde{\psi} \sim \frac{\Gamma (c) \Gamma \left(a_--a_+\right)}{\Gamma \left(a_-\right) \Gamma \left(c-a_+\right)}z^{i \left(q-2 \tilde{\omega }\right) r_+-a_+}+\frac{\Gamma (c) \Gamma \left(a_+-a_-\right)
   }{\Gamma \left(a_+\right)\Gamma \left(c-a_-\right)}z^{i \left(q-2 \tilde{\omega }\right) r_+-a_-}\,.
\end{equation}
If $m^2_{\mathrm{eff}}L_{\mathrm{AdS}_{2}}^2+\frac{1}{4}\geq0$, we expect the corresponding quasinormal mode to be localized near the horizon, and so we should demand the growing mode to vanish, \emph{i.e.} the second term in the expansion \eqref{appNHlargez} should vanish. This is achieved if we set
\begin{equation}
a_+=-n\,,\quad\text{with}\quad n\in \mathbb{N}^0\Rightarrow \tilde{\omega}=\frac{q}{2}-\frac{i}{2\,r_+}\left(\frac{1}{2}+n+\sqrt{m^2_{\mathrm{eff}}L_{\mathrm{AdS}_{2}}^2+\frac{1}{4}}\right)\,,
\end{equation}
and thus
\begin{equation}
\omega = \frac{1-\mu^2}{2}\left[q-\frac{i}{r_+}\left(\frac{1}{2}+n+\sqrt{m^2_{\mathrm{eff}}L_{\mathrm{AdS}_{2}}^2+\frac{1}{4}}\right)\right]\,.
\label{eq:nearhoomega}
\end{equation}
We expect to see a matching of the expression above with our numerical results so long as $m^2_{\mathrm{eff}}L_{\mathrm{AdS}_{2}}^2+\frac{1}{4}\geq 0$ (\emph{i.e} if the near horizon $\mathrm{AdS}_2$ BF bound condition holds). We will confirm this is the case in Fig.~\ref{fig:nearextremal} and associated discussion in the end of Sec.~\ref{sec:A2}.

Note that the analysis of this subsection predicts stability when the condition $m^2_{\mathrm{eff}}L_{\mathrm{AdS}_{2}}^2+\frac{1}{4}\geq0$ holds, but says nothing about the stability of the system in the complementary regime, \emph{i.e.} when $m^2_{\mathrm{eff}}L_{\mathrm{AdS}_{2}}^2+\frac{1}{4}<0$. As reported in Sec.~\ref{sec:linear} and in the next subsections of this Appendix, we actually find that the system is unstable in the latter regime where the $\mathrm{AdS}_2$ BF bound condition is violated. 
\subsection{Numerical results for non-extremal black holes}\label{sec:A2}

We now proceed to compute numerically the frequency spectrum of the eigenvalue problem~\eqref{eq:linear}. Determining the decay or growth rates of the modes in the full parameter space will definitely establish in which conditions the system is stable/unstable and, in particular, confirm the analytical predictions found in sections \ref{sec:linearNH} and \ref{sec:A1}. 

A Frobenius analysis of~\eqref{eq:linear} near the black hole event horizon $r=r_+$, reveals that $\tilde{\psi}$ admits the following expansion
\begin{equation}
\tilde{\psi} = \left(1-\frac{r_+}{r}\right)^{i\,\frac{r_+\,\omega}{1-\mu^2}}C_+\Big\{1+\mathcal{O}[(r-r_+)]\Big\} + \left(1-\frac{r_+}{r}\right)^{-i\,\frac{r_+\,\omega}{1-\mu^2}}C_-\Big\{1+\mathcal{O}[(r-r_+)]\Big\}.
\label{eq:frobho}
\end{equation}
We wish to impose regularity across the black hole's future event horizon, that is to say, we want $\tilde{\psi}$ to be regular in \emph{ingoing Eddington-Finkelstein} coordinates. This procedure then implies that physically acceptable solutions must have $C_+=0$.

Near spatial infinity, \emph{i.e.} as $r\to+\infty$, we find
\begin{multline}
\tilde{\psi} = A_+ e^{\sqrt{m^2-(\omega+q\mu)^2}\,r}\left(\frac{r_+}{r}\right)^{1+\eta_+}\left[1+\mathcal{O}\left(\frac{r_+}{r}\right)\right]
\\
+A_- e^{-\sqrt{m^2-(\omega+q\mu)^2}\,r}\left(\frac{r_+}{r}\right)^{1+\eta_-}\left[1+\mathcal{O}\left(\frac{r_+}{r}\right)\right]
\end{multline}
with
\begin{equation}
\eta_{\pm}=\pm\left[(\omega +q\mu)\frac{(\omega +q\mu)(\mu M-Q)+r_+\mu \,\omega}{\mu\sqrt{m^2-(\omega +q\mu)^2}}-M\,\sqrt{m^2-(\omega +q\mu)^2}\right]\,.
\end{equation}
Since we are interested in finite energy excitations we demand that physically acceptable solutions decay appropriately as $r\to+\infty$. On can show that the term proportional to $A_+$ diverges at large $r$, so we take $A_+=0$ as our asymptotic boundary condition.

The boundary conditions above suggest we should not work directly with $\tilde{\psi}$, but instead define a new function $\hat{\psi}$ through the relation
\begin{equation}
\tilde{\psi} \equiv  e^{-\sqrt{m^2-(\omega+q\mu)^2}\,r}\left(\frac{r_+}{r}\right)^{1+\eta_-}\left(1-\frac{r_+}{r}\right)^{-i\,\frac{r_+\,\omega}{1-\mu^2}}\hat{\psi}\,.
\label{eq:off2}
\end{equation}
Unlike $\tilde{\psi}$, $\hat{\psi}$ is now smooth for the desired boundary conditions at both $r=r_+$ and as $r\to+\infty$.

Numerically, it is hard to work with infinite domains so we introduce a compact coordinate $y$ given by
\begin{equation}
r=\frac{r_+}{1-y}\,,
\label{eq:ycoord2}
\end{equation}
with the horizon located at $y=0$ and asymptotic infinity at $y=1$. The boundary conditions for $\hat{\psi}$ are then found by demanding $\hat{\psi}$ to have a regular Taylor expansion at $y=0$  and $y=1$. This procedure yields rather cumbersome Robin boundary conditions at $y=0$ and $y=1$ which we do not present here.

The way forward is now clear: for each value of $q/m$, $\alpha$, $m \,r_+$ and $\mu$ (or alternatively, $q/m$, $\alpha$, $m Q$ and $m M$), we determine $\omega/m$ via a generalised eigenvalue problem. In general, $\omega$ will be complex. When $\mathrm{Im}\, \omega<0$ we have a decaying mode or \emph{quasinormal mode}, whereas when $\mathrm{Im}\, \omega>0$ we have an instability. At $\mathrm{Im}\, \omega=0$ we then have the instability onset.

\begin{figure}[b!]
    \centering
    \vspace{1cm}
    \includegraphics[width=0.85\textwidth]{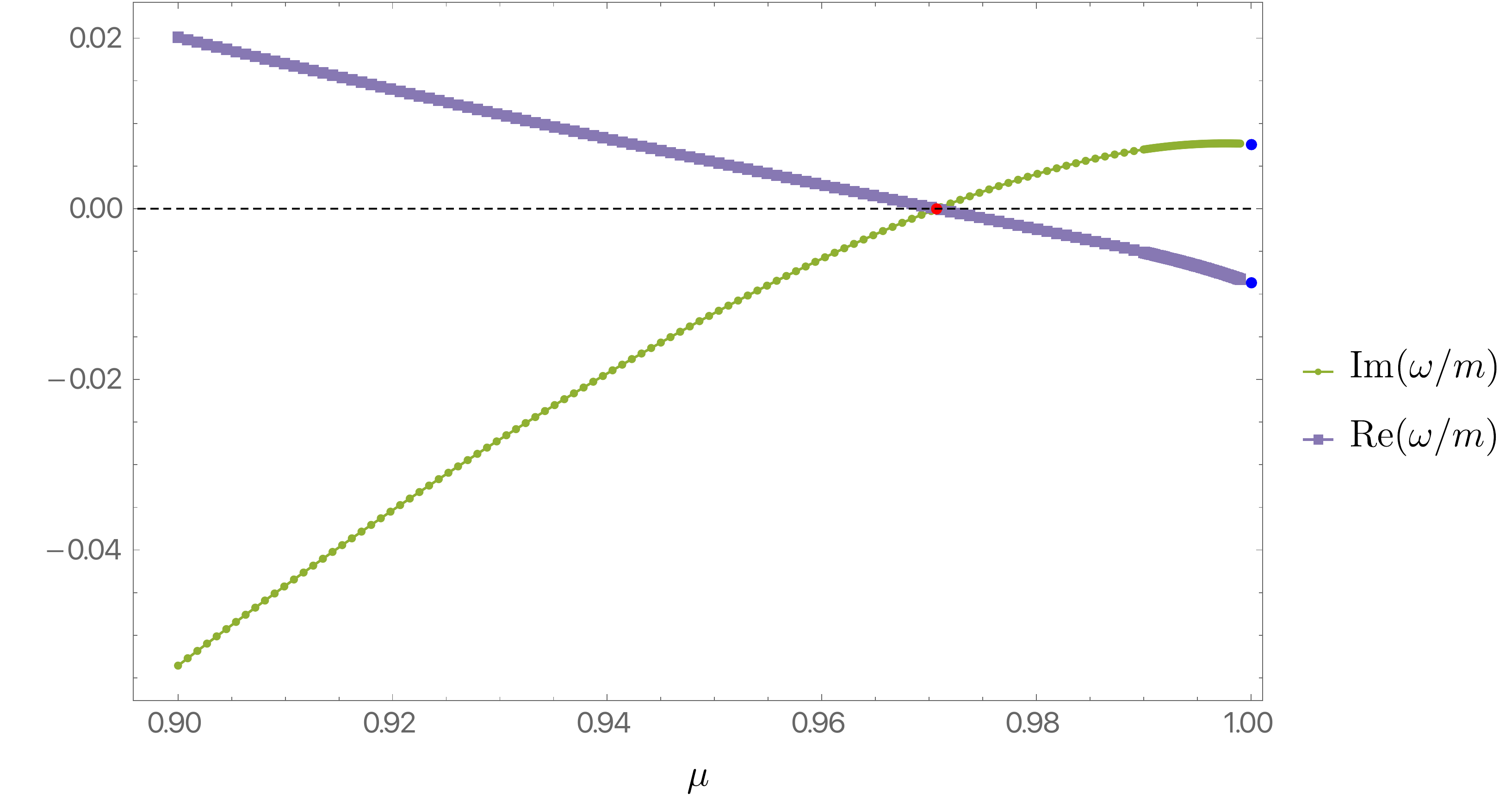}
    \caption{Imaginary (green disks) and real (purple squares) parts of $\omega$ as a function of $\mu$ for fixed $\alpha=1$, $q/m=0.5$ and $m \, r_+ =1$. The red point with $\mathrm{Im}\, \omega=0$ and $\mathrm{Re}\, \omega=0$ is the onset of the instability and the merger point between RN and hairy black holes (obtained using the time dependent analysis of Sec.~\ref{sec:A2} and, independently, the static analysis of Sec.~\ref{sec:linearOnset}). The blue points at $\mu=1$ correspond to the frequency computed directly at extremality using the method outlined in section \ref{sec:A3}.}
       \label{fig:example}
\end{figure}
Let us imagine that we fix $\alpha$, $q/m$ and $m \,r_+$. We are left with a line in moduli space parametrised by $\mu$. As we increase $\mu$ towards extremality, we expect $\mathrm{Im}\, \omega$ to start off negative, and become positive near extremality. At some point, we will find a value $\mu=\mu_c$ at which $\mathrm{Im}\, \omega=0$. One can show that, given our boundary conditions, at this point we must also have $\mathrm{Re}\, \omega=0$. 
We are thus left with a particular value of $\mu=\mu_c$ at which the instability sets in. This instability onset is also the point where hairy black holes smoothly join the RN black hole in a phase diagram of solutions of the theory.
As we vary $m \, r_+$ at fixed $\alpha$ and $q/m$, we trace a line in the moduli space $(m M,m Q)$ which we coin the \emph{merger line}. An example of this procedure is presented in Fig.~\ref{fig:example}, which was obtained by taking $\alpha=1$, $q/m=0.5$ and $r_+m = 1$. We can indeed confirm that before we reach extremality (around $\mu\sim 0.970729$), $\mathrm{Im}\, \omega$ becomes positive, signaling an exponentially growing instability.

There is a complementary way to determine the merger line, as described in Sec.~\ref{sec:linearOnset}. We start by setting $\omega=0$ in all the above expressions, and regard the problem of finding a smooth, perturbative, hairy solution as a generalised eigenvalue problem in $\mu$ (at fixed $m\,r_+$, $\alpha$ and $q/m$). Both methods yield the same critical onset/merger value for $\mu_c$, which is reassuring.

The results in Fig.~\ref{fig:example} are concordant with our expectations. In particular, for $\alpha=1$, $q/m=0.5$ and $m\,r_+ =1$, we have $m^2_{\mathrm{eff}}L_{\mathrm{AdS}_{2}}^2=-1.0625<-0.25$: the $\mathrm{AdS}_2$ BF bound is violated and thus we expect an instability before we reach extremality (see analysis of Sec.~\ref{sec:linearNH}). 

\begin{figure}[b!]
    \centering
    \vspace{1cm}
    \includegraphics[width=\textwidth]{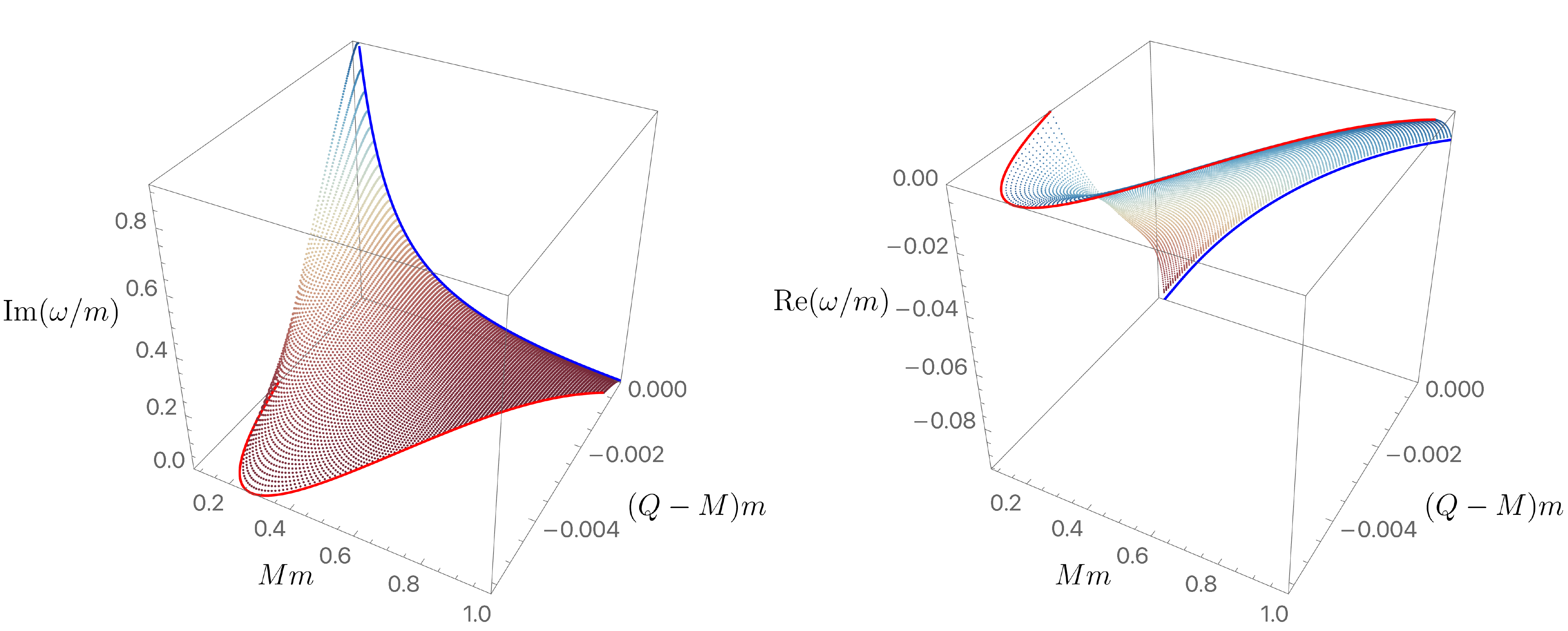}
    \caption{Imaginary part (left panel) and real part (right panel) of $\omega$ as a function of $m M$ and $(Q-M)m$ for fixed $\alpha=1$, $q/m=0.5$. The onset line, depicted in red, was computed directly with $\omega=0$ (see Sec.~\ref{sec:linearOnset}), whereas the blue line at extremality, where $M=Q$, was computed using the method outlined in section \ref{sec:A3}.}
       \label{fig:3D}
\end{figure}
In Fig.~\ref{fig:3D} we plot $\mathrm{Im}(\omega/m)$ (left panel) and $\mathrm{Re}(\omega/m)$ (right panel) as a function of $m M$ and $(Q-M)m$ for fixed $\alpha=1$ and $q/m=0.5$. For clarity, we only show the region in moduli space where the RN black hole is unstable. The growth rate appears to grow as we approach extremality, at fixed $m M$, as expected. We also plot in Fig.~\ref{fig:3D} the merger line (where $\mathrm{Im}\, \omega=0$). Finally, we add to this plot a blue line computed directly at extremality (see \ref{sec:A3}). The agreement between our non-extremal data and the data computed at extremality is reassuring. We have repeated this plot for several values of $\alpha$ and $q/m$, and the overall qualitative picture appears the same.


\begin{figure}[t!]
    \centering
    \vspace{1cm}
    \includegraphics[width=\textwidth]{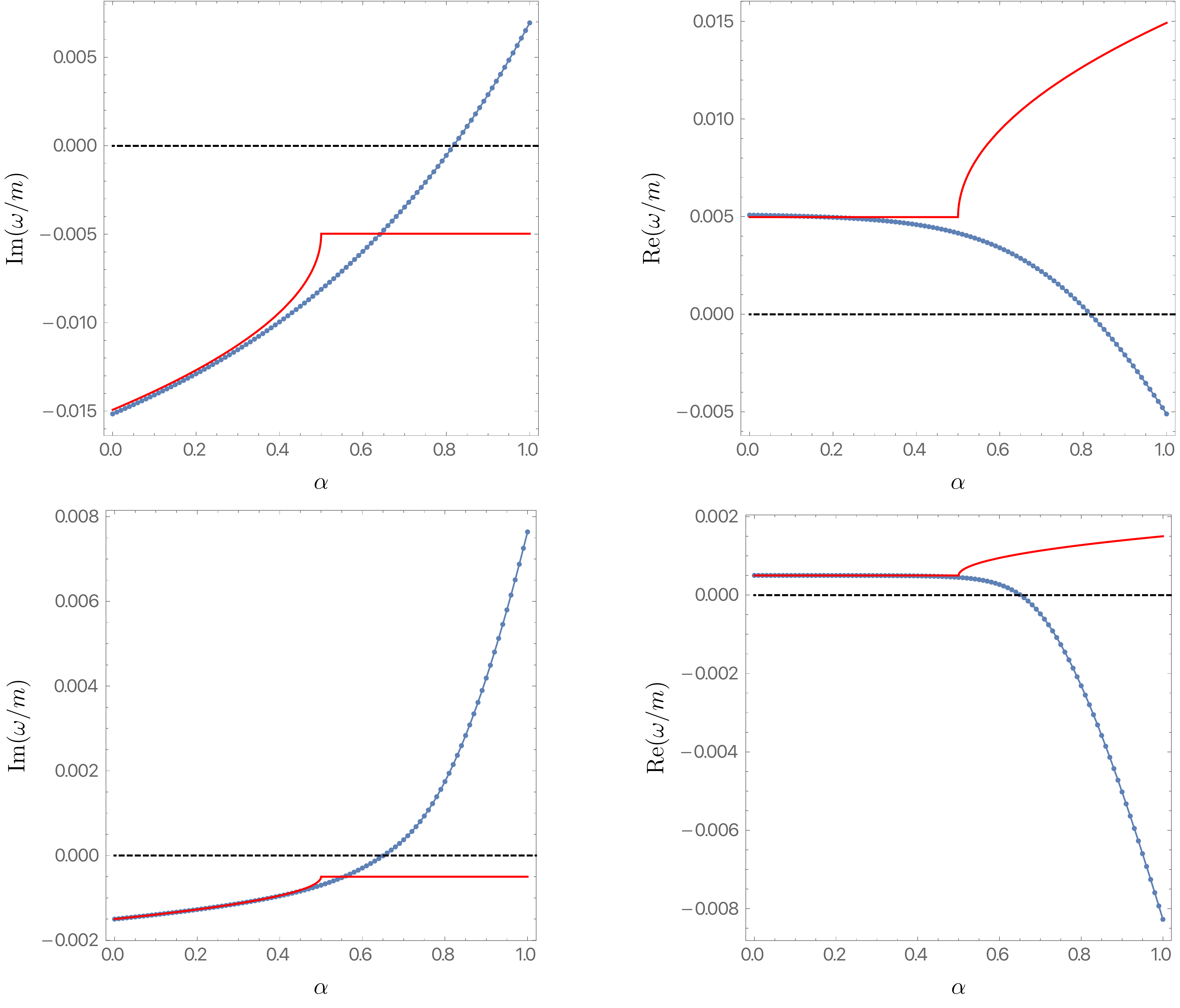}
    \caption{Imaginary (left column) and real (right column) parts of $\omega$ as a function of $\alpha$ for fixed $q/m=0.5$ and $m \, r_+ =1$. The top row has $\mu=0.99$, whereas the bottom row has $\mu=0.999$. The disks represent numerical data, whereas solid red lines yield the analytic prediction given in \eqref{eq:nearhoomega}. The agreement between the analytic predictions and exact numerical data is excellent when $\alpha<0.5$, \emph{i.e.} $m^2_{\mathrm{eff}}L_{\mathrm{AdS}_{2}}^2+1/4\geq 0$.}
       \label{fig:nearextremal}
\end{figure}
Using our numerical results, we can also test whether \eqref{eq:nearhoomega} provides a good approximation when the $\mathrm{AdS}_2$ BF bound is preserved, \emph{i.e.} when $m^2_{\mathrm{eff}}L_{\mathrm{AdS}_{2}}^2+1/4\geq 0$. In Fig.~\ref{fig:nearextremal} we have $\mu = 0.99$ on the top two plots, and $\mu=0.999$ on the bottom two plots. In both cases we have $m \, r_+ = 1$. On the left column we plot the imaginary part, while on the right column we plot the real part. In all cases, the horizontal axis is given by the Maxwell-scalar coupling $\alpha$. The blue disks in each figure label the exact numerical data, while the red solid lines give the prediction \eqref{eq:nearhoomega}. As anticipated, \eqref{eq:nearhoomega} only yields a good approximation when $m^2_{\mathrm{eff}}L_{\mathrm{AdS}_{2}}^2+1/4\geq0$, which for the parameters used in generating Fig.~\ref{fig:nearextremal} corresponds to $\alpha\leq 0.5$. Furthermore, the closer we get to extremality, \emph{i.e.} the closer $\mu$ gets to unity, the better the approximation should become. Indeed, the red solid lines all appear to get closer to the numerical data (represented by the blue disks) on the lower row of plots.

One might wonder, however, whether the bound $m^2_{\mathrm{eff}}L_{\mathrm{AdS}_{2}}^2+1/4\geq0$ really is sharp. This is a question that can only be answered by working directly at extremality, since extremal black holes are the ones we expect to be most unstable. We address this question in the next subsection. 

\subsection{Numerical results for extremal black holes\label{sec:A3}}
Computing the instability growth rate directly at extremality requires some care. The reason being that the Frobenius expansion near the black hole event horizon no longer takes the form given in \eqref{eq:frobho}. Instead, one now finds that a Frobenius expansion at this degenerate horizon yields
\begin{multline}
\tilde{\psi}= C_+\,e^{\frac{i \omega r_+}{1-\frac{r_+}{r}}}\left(1-\frac{r_+}{r}\right)^{-i r_+(q+2\omega)}\left\{1+\mathcal{O}\left[(r-r_+)\right]\right\}+
\\
C_-\,e^{-\frac{i \omega r_+}{1-\frac{r_+}{r}}}\left(1-\frac{r_+}{r}\right)^{i r_+(q+2\omega)}\left\{1+\mathcal{O}\left[(r-r_+)\right]\right\}\,.
\end{multline}
Regularity at the future event horizon now demands $C_-=0$, which in turn suggests that we change \eqref{eq:off2} into
\begin{equation}
\tilde{\psi} \equiv  e^{-\sqrt{m^2-(\omega+q)^2}\,r+\frac{i \omega r_+}{1-\frac{r_+}{r}}}\left(\frac{r_+}{r}\right)^{1+\eta_-}\left(1-\frac{r_+}{r}\right)^{-i r_+(q+2\omega)}\hat{\psi}\,,
\end{equation}
where we should set $\mu=1$ when computing $\eta_-$. We again use the $y$ coordinate introduced in \eqref{eq:ycoord2}. The agreement between our extremal code and the non-extremal code can be observed in Fig.~\ref{fig:example}, where the blue disks (exactly at $\mu=1$) were computed directly at extremality and appear to be a natural continuation of the near extremal curves (green and purple disks). The agreement is also clear in Fig.~\ref{fig:3D} where the non-extremal surface approaches the blue line ($Q-M=0$) computed directly at extremality.  

\begin{figure}[h!]
    \centering
    \vspace{1cm}
    \includegraphics[width=0.65\textwidth]{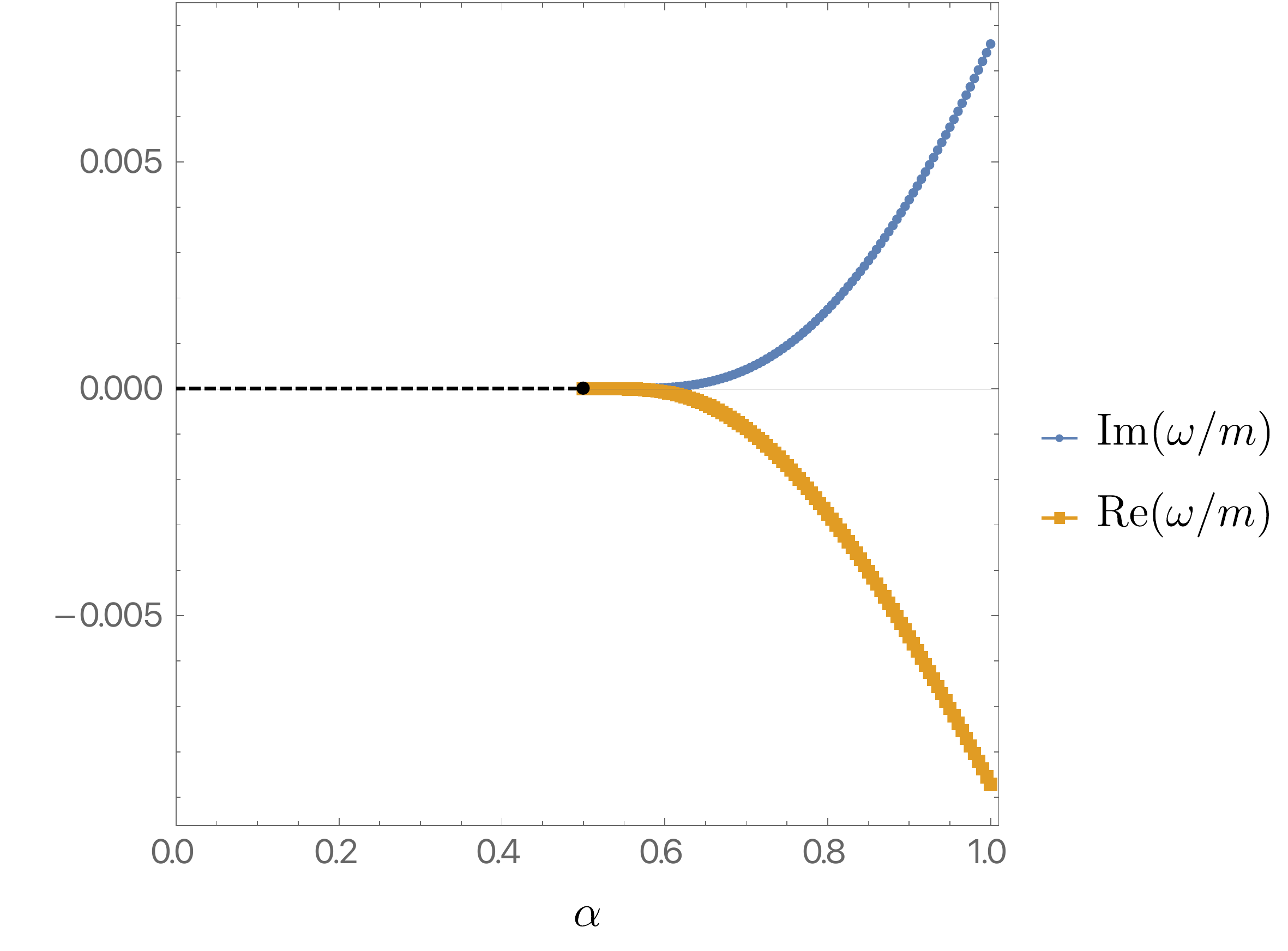}
    \caption{Imaginary (blue disks) and real (orange squares) parts of $\omega$ as a function of $\alpha$ for fixed $q/m=0.5$ and $m \, r_+ =1$. The frequency vanish for $\alpha\leq 0.5$ (black disk), in accordance to the $\mathrm{AdS}_2$ BF bound condition given by \eqref{eq:bound} and the complementary analysis of Sec.~\ref{sec:A1}.}
       \label{fig:extremalfrequency}
\end{figure}

To conclude this appendix, in Fig.~\ref{fig:extremalfrequency} we show the frequency computed directly at extremality as a function of the Maxwell-scalar coupling $\alpha$, for fixed $m \, r_+=1$ and $q/m=0.5$. The $\mathrm{AdS}_2$ BF bound violation condition \eqref{eq:bound} now predicts that an instability should only exist if $\alpha>1/2$. This is precisely what we observe in Fig.~\ref{fig:extremalfrequency}. Note also that, in the complementary regime where the $\mathrm{AdS}_2$ BF bound is not violated, according to \eqref{eq:nearhoomega} one should have $\omega=0$ for $\alpha<1/2$, which is again what we find in Fig.~\ref{fig:extremalfrequency}. We have repeated this for many other values of $\alpha$, $q/m$ and $m \, r_+$ and always found \eqref{eq:bound} to appear sharp. In fact, one can go further and use a similar technique to the one appearing in section 6.4 of \cite{Dias:2010ma} and construct initial data with negative energy, thus showing instability.

\bibliographystyle{jhep}
	\cleardoublepage

\renewcommand*{\bibname}{References}

\bibliography{kasner}

\end{document}